\documentclass{aa}
\usepackage[varg]{txfonts}
\usepackage{natbib}
\usepackage{color}
\usepackage{rotating}
\usepackage{graphicx}
\usepackage{textcomp}
\usepackage{orcidlink}
\bibpunct{(}{)}{;}{a}{}{,}

\begin{document}

\title{The ALMA Frontier Fields Survey}
\subtitle{VI. Lensing-corrected 1.1\,mm number counts in Abell 2744, MACSJ0416.1-2403, MACSJ1149.5+2223, Abell 370 and Abell S1063}

\author{A. M. Mu\~noz Arancibia\inst{1,2}\orcidlink{0000-0002-8722-516X} \and J. Gonz\'alez-L\'opez\inst{3,4}\orcidlink{0000-0003-3926-1411} \and E. Ibar\inst{1} \and F. E. Bauer\inst{5,2,6}\orcidlink{0000-0002-8686-8737} \and T. Anguita\inst{7,2}\orcidlink{0000-0003-0930-5815} \and M. Aravena\inst{4}\orcidlink{0000-0002-6290-3198} \and R. Demarco\inst{8}\orcidlink{0000-0003-3921-2177} \and R. Kneissl\inst{9,10}\orcidlink{0000-0002-5580-006X} \and A. M. Koekemoer\inst{11}\orcidlink{0000-0002-6610-2048} \and P. Troncoso-Iribarren\inst{12} \and A. Zitrin\inst{13}\orcidlink{0000-0002-0350-4488}}

\institute{Instituto de F\'isica y Astronom\'ia, Universidad de Valpara\'iso, Av. Gran Breta\~na 1111, Valpara\'iso, Chile\\e-mail: \texttt{amma.19@gmail.com}
\and Millennium Institute of Astrophysics, Nuncio Monse\~nor S\'otero Sanz 100, Providencia, Santiago, Chile
\and Las Campanas Observatory, Carnegie Institution of Washington, Casilla 601, La Serena, Chile
\and N\'ucleo de Astronom\'ia de la Facultad de Ingenier\'ia y Ciencias, Universidad Diego Portales, Av. Ej\'ercito Libertador 441, Santiago, Chile
\and Instituto de Astrof\'isica y Centro de Astroingenier\'ia, Facultad de F\'isica, Pontificia Universidad Cat\'olica de Chile, Casilla 306, Santiago 22, Chile
\and Space Science Institute, 4750 Walnut Street, Suite 205, Boulder, CO 80301, USA
\and Departamento de Ciencias Fisicas, Universidad Andres Bello, Fernandez Concha 700, Las Condes, Santiago, Chile
\and Departamento de Astronom\'ia, Facultad de Ciencias F\'isicas y Matem\'aticas, Universidad de Concepci\'on, Concepci\'on, Chile
\and Joint ALMA Observatory, Alonso de C\'ordova 3107, Vitacura, Santiago, Chile
\and European Southern Observatory, Alonso de C\'ordova 3107, Vitacura, Casilla, 19001 Santiago, Chile
\and Space Telescope Science Institute, 3700 San Martin Dr., Baltimore, MD 21218, USA
\and Escuela de Ingenier\'ia, Universidad Central de Chile, Avenida Francisco de Aguirre 0405, 171-0614, La Serena, Coquimbo, Chile
\and Physics Department, Ben-Gurion University of the Negev, PO Box 653, Be’er-Sheva 8410501, Israel
}

\date{}

\abstract{Probing the faint end of the number counts at mm wavelengths is important in order to identify the origin of the extragalactic background light in this regime. Aided by strong gravitational lensing, ALMA observations towards massive galaxy clusters have opened a window to probe this origin, allowing to resolve sub-mJy dusty star-forming galaxies.}
{We aim to derive number counts at $1.1\,\textmd{mm}$ down to flux densities fainter than $0.1\,\textmd{mJy}$, based on ALMA observations towards five \textit{Hubble} Frontier Fields (FF) galaxy clusters, following a statistical approach to correct for lensing effects.}
{We created a source catalog that includes ALMA $1.1\,\textmd{mm}$ continuum detections around two new FF galaxy clusters, together with the sources previously detected around three FF galaxy clusters, making a total of 29 detected sources down to a $4.5\sigma$ significance. ALMA 1.1mm mosaics used for our source extraction covered the inner $\approx2'\times2'$ FF regions, reached rms depths of $\approx55-71\,\mu\textmd{Jy}\,\textmd{beam}^{-1}$, and had synthesized beam sizes between $\approx0\farcs5-1\farcs5$ (natural weighting). We derived source intrinsic flux densities using public lensing models. We folded the uncertainties in both magnifications and source redshifts into the number counts through Monte Carlo simulations.}
{Using the combination of all cluster fields, we derive cumulative number counts over two orders of magnitude down to $\approx0.01\,\textmd{mJy}$ after correction for lensing effects. Cosmic variance estimates are all exceeded by uncertainties in our median combined cumulative counts that come from both our Monte Carlo simulations and Poisson statistics. Our number counts agree at a $1\sigma$ level with our previous estimates using ALMA observations of the first three FFs, exhibiting a similar flattening at faint flux densities. They are also consistent to $1\sigma$ with most of recent ALMA estimates and galaxy evolution models. However, below $\approx0.1\,\textmd{mJy}$, our cumulative number counts are lower by $\approx0.4\,\textmd{dex}$ compared to two deep ALMA studies (namely one that probes several blank fields plus one lensed galaxy cluster, and the initial ALMA Spectroscopic Survey in the Hubble Ultra Deep Field, ASPECS-Pilot), while remaining consistent with the ASPECS Large Program (ASPECS-LP) within $1\sigma$. Importantly, the flattening found for our cumulative counts at $\lesssim0.1\,\textmd{mJy}$ also extends further to $\approx0.01\,\textmd{mJy}$, i.e., $\approx0.4\,\textmd{dex}$ fainter than ASPECS-LP, and remains in agreement with extrapolations of their number counts down to this flux limit. We find a median contribution to the extragalactic background light (EBL) of $14_{-8}^{+12}\,\textmd{Jy}\,\textmd{deg}^{-2}$ resolved in our demagnified sources down to $\approx0.01\,\textmd{mJy}$, representing $75-86\%$ of Planck-derived extragalactic EBL estimates at $1.1\,\textmd{mm}$.}
{We estimate cumulative $1.1\,\textmd{mm}$ number counts down to $\approx0.01\,\textmd{mJy}$, along the line of sight of five galaxy clusters that benefit from having rich deep multiwavelength data. They bring further support in line of the flattening of the number counts reported previously by us and ASPECS-LP, which has been interpreted by a recent galaxy evolution model as a measurement of the "knee" of the infrared luminosity function at high redshift. Our estimates of the contribution to the EBL associated with $1.1\,\textmd{mm}$ galaxies in the FFs suggest that we may be resolving most of the EBL at this wavelength down to $\approx0.01\,\textmd{mJy}$.}


\maketitle

\section{Introduction}

Observations of the extragalactic background light (EBL) across the whole electromagnetic spectrum have shown that the integrated intensity in the far infrared (FIR) to millimeter (mm) regime is comparable to that of the optical portion \citep{Puget1996,Fixsen1998,Dole2006,Cooray2016}. Attempting to resolve this cosmic infrared background into distinct sources led to the discovery of submillimeter galaxies (e.g., \citealt{Smail1997,Hughes1998}), which appear to be at high redshift and harbor dust-obscured star formation \citep{Blain2002}. Since these first detections, substantial progress has been made characterizing the also-called dusty star-forming galaxies (DSFGs; see \citealt{Casey2014}, for a review). The high spatial resolution and sensitivity of the Atacama Large Millimeter/submillimeter Array (ALMA) is allowing to probe DSFGs in exquisite detail, giving more insights into the nature of star-forming galaxies at high redshift. Importantly, deep ALMA observations may help to resolve the faint end of the dusty galaxy population (i.e., sources having mm flux densities below $0.1\,\textmd{mJy}$), thus giving more clues to the origin of the EBL at mm wavelengths. While currently there are plenty of brighter DSFG detections (via single-dish and/or interferometric observations; see e.g., \citealt{Casey2014,Hodge2020}), only a few studies have uncovered the faint-end population \citep{Fujimoto2016,Aravena2016,Gonzalez-Lopez2017,Gonzalez-Lopez2020}.

One of these projects is the ALMA Frontier Fields Survey (PI: F. Bauer). Introduced in \citet[hereafter Paper I]{Gonzalez-Lopez2017}, it looks for DSFGs using deep ALMA Band 6 observations towards \textit{Hubble} Frontier Fields (FF) galaxy clusters \citep{Lotz2017}. This ALMA survey benefits from the unique ALMA capabilities and the strong gravitational lensing power of massive galaxy clusters, allowing to detect (if magnified) sources beyond the confusion limit (e.g., rms depths in the order of tens of $\mu\textmd{Jy}\,\textmd{beam}^{-1}$ for current deep ALMA maps at $1.1\,\textmd{mm}$). It also exploits the public availability of deep multiwavelength data (including HST, \textit{Spitzer} and VLA) and several detailed mass models for each galaxy cluster, both of which are at the core of the FF legacy project. The ALMA FF Survey has successfully found and characterized DSFGs both in the continuum (see Paper I) and emission lines \citep[hereafter Paper III]{Gonzalez-Lopez2017b}. It has also included a multiwavelength photometric analysis of a sample of ALMA detections \citep[hereafter Paper II]{Laporte2017} and an ALMA stacking of Lyman-break galaxies in the FFs \citep{Carvajal2020}.

In \citet[hereafter Paper IV]{MunozArancibia2018}, we reported $1.1\,\textmd{mm}$ number counts for the first three FF galaxy clusters (see also \citealt[hereafter Paper IV Corrigendum]{MunozArancibia2019}), following a detailed treatment of the uncertainties that come from the lensing models. Based on 19 ALMA $1.1\,\textmd{mm}$ continuum detections, we derived counts spanning around two orders of magnitude in demagnified flux density, down to $\approx0.01\,\textmd{mJy}$. Although these counts were consistent at a $1\sigma$ level with most of deep ALMA observations reported by that time, we found that below $\approx0.1\,\textmd{mJy}$ our cumulative counts were lower by $\approx0.5\,\textmd{dex}$, suggesting a flattening in the number counts. Encouragingly, this flattening was further confirmed independently by the ALMA Spectroscopic Survey in the Hubble Ultra Deep Field Large Program (ASPECS-LP, \citealt{Gonzalez-Lopez2020,Aravena2020}). Based on the deepest $1.2\,\textmd{mm}$ data to date in a contiguous area over the sky, \citet{Gonzalez-Lopez2020} derived number counts down to $\approx0.03\,\textmd{mJy}$, noticing a flattening in the cumulative counts below $0.3\,\textmd{mJy}$. In this work, we expand our census of the surface density of DSFGs detected at $1.1\,\textmd{mm}$ in the FFs, through galaxy number counts that include the remaining galaxy clusters that form our survey. We aim to derive number counts again down to flux densities fainter than $0.1\,\textmd{mJy}$, bolstering our previous results with better statistics.

This paper is organized as follows. Section \ref{sect_data} presents the ALMA observations, redshift estimates and lensing models here used. Section \ref{sect_method} describes the procedure used to derive lensing-corrected number counts. Section \ref{sect_results} reports our $1.1\,\textmd{mm}$ counts and compares them with our previous work, recent literature data using ALMA, and galaxy evolution models. It also presents our estimates of cosmic variance and contribution to the EBL based on our number counts. Section \ref{sect_concl} presents a summary of our findings. We adopt a flat $\Lambda$CDM cosmology with parameters $H_0=70\,\textmd{km}\,\textmd{s}^{-1}\textmd{Mpc}^{-1}$, $\Omega_m=0.3$ and $\Omega_{\Lambda}=0.7$.

\section{Data}\label{sect_data}

\subsection{Observations with ALMA}\label{sect_obs}

\begin{table*}
\begin{center}
\caption{ALMA mosaic properties of galaxy clusters.}
\begin{tabular}{ccccccc}
\hline \hline
Cluster name & $z$ & $\textmd{RA}_{\textmd{J2000}}$ & $\textmd{Dec}_{\textmd{J2000}}$ & $b_{\textmd{max}}\times b_{\textmd{min}}$ & $b_{\textmd{PA}}$ & rms\\
&& (hh:mm:ss.ss) & ($\pm$dd:mm:ss.ss) & $(''\times'')$ & (\textdegree) & $(\mu\textmd{Jy}\,\textmd{beam}^{-1})$\\
(1) & (2) & (3) & (4) & (5) & (6) & (7) \\
\hline
Abell 2744 & 0.308 & 00:14:21.2 & $-$30:23:50.1 & $0.63\times0.49$ & 86.14 & 55\\
MACSJ0416.1-2403 & 0.396 & 04:16:08.9 & $-$24:04:28.7 & $1.52\times0.85$ & $-$85.13 & 59\\
MACSJ1149.5+2223 & 0.543 & 11:49:36.3 & +22:23:58.1 & $1.22\times1.08$ & $-$43.46 & 71\\
Abell 370 & 0.375 & 02:39:52.9 & $-$01:34:36.5 & $1.25\times1.00$ & 89.30 & 62\\
Abell S1063 & 0.348 & 22:48:44.4 & $-$44:31:48.5 & $0.96\times0.79$ & $-$80.33 & 67\\
\hline
\end{tabular}
\tablefoot{Column 1: Cluster name. Column 2: Cluster redshift. Columns 3, 4: Central J2000 position of mosaic. Column 5: Major and minor axes of synthesized beam. Column 6: Position angle of synthesized beam. Column 7: Lowest rms (highest sensitivity) achieved in mosaic.}
\label{tab_clus}
\end{center}
\end{table*}

\begin{figure*}
\centering
\resizebox{0.88\hsize}{!}{\includegraphics{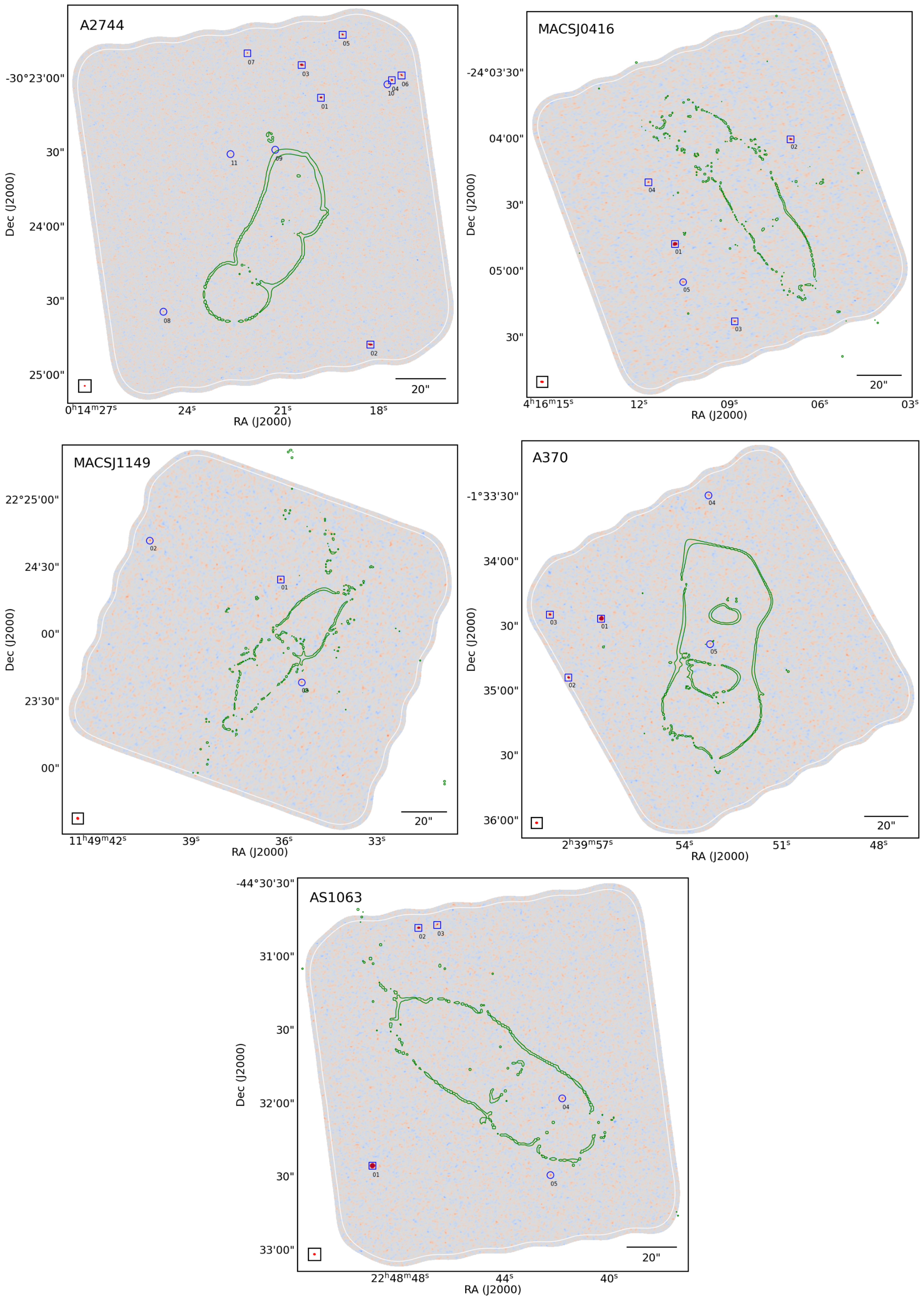}}
\caption{ALMA continuum maps at $1.1\,\textmd{mm}$ for five Frontier Fields clusters, made with natural weighting. Images are displayed without PB sensitivity corrections. Color scale goes from blue to red, corresponding to $-5\sigma$ to $5\sigma$. A white contour in each map encloses the region with $\textmd{PB}>0.3$, that is, where the PB sensitivity is at least 30\% of the peak sensitivity. Source positions are displayed in blue as squares (high-significance) or circles (low-significance). A small box in the bottom left corner of each panel shows the synthesized beam (see parameters in Table \ref{tab_clus}). Green lines indicate the critical curves (i.e. infinite magnification) for a $z=2$ source plane in the Diego v4.1 lens model (see references in Table \ref{tab_models}).}
\label{fig_maps}
\end{figure*}

\begin{table*}
\begin{center}
\caption{Continuum detections at $\textmd{S/N}\geq4.5$.}
\begin{tabular}{cccccccc}
\hline \hline
ID & $\textmd{RA}_{\textmd{J2000}}$ & $\textmd{Dec}_{\textmd{J2000}}$ & S/N & $S_{1.1\,\textmd{mm,peak}}$ & $S_{1.1\,\textmd{mm,}uv\textmd{-fit}}$ & $r_s$ & $z$\\
& (hh:mm:ss.ss) & ($\pm$dd:mm:ss.ss) & & $(\textmd{mJy}\,\textmd{beam}^{-1})$ & $(\textmd{mJy})$ & $('')$ &\\
(1) & (2) & (3) & (4) & (5) & (6) & (7) & (8)\\
\hline
A2744-ID01\tablefootmark{a} & 00:14:19.80 & $-$30:23:07.66 & 25.9 & $1.433\pm0.056$ & $1.570\pm0.073$ & $0.05\pm0.01$ & 2.9\tablefootmark{c}\\
A2744-ID02\tablefootmark{a} & 00:14:18.25 & $-$30:24:47.47 & 14.4 & $1.292\pm0.091$ & $2.816\pm0.229$ & $0.23\pm0.04$ & 2.585\tablefootmark{d}\\
A2744-ID03\tablefootmark{a} & 00:14:20.40 & $-$30:22:54.42 & 13.9 & $0.798\pm0.058$ & $1.589\pm0.125$ & $0.26\pm0.03$ & 3.058\tablefootmark{d}\\
A2744-ID04\tablefootmark{a} & 00:14:17.58 & $-$30:23:00.56 & 13.8 & $0.932\pm0.068$ & $1.009\pm0.074$ & $0.06\pm0.02$ & 1.498\tablefootmark{d}\\
A2744-ID05\tablefootmark{a} & 00:14:19.12 & $-$30:22:42.20 & 7.7 & $0.655\pm0.086$ & $1.113\pm0.135$ & $0.19\pm0.05$ & 2.409\tablefootmark{d}\\
A2744-ID06\tablefootmark{a} & 00:14:17.28 & $-$30:22:58.60 & 6.5 & $0.574\pm0.089$ & $1.283\pm0.241$ & $0.26\pm0.08$ & $2.08_{-0.08}^{+0.13}$\tablefootmark{e}\\
A2744-ID07\tablefootmark{a} & 00:14:22.10 & $-$30:22:49.67 & 6.2 & $0.455\pm0.074$ & $0.539\pm0.082$ & $0.07\pm0.04$ & 2.644\tablefootmark{d}\\
A2744-ID08\tablefootmark{b} & 00:14:24.73 & $-$30:24:34.20 & 4.8 & $0.270\pm0.056$ & \dots  & \dots & \dots\\
A2744-ID09\tablefootmark{b} & 00:14:21.23 & $-$30:23:28.70 & 4.7 & $0.256\pm0.055$ & \dots  & \dots & \dots\\
A2744-ID10\tablefootmark{b} & 00:14:17.72 & $-$30:23:02.25 & 4.5 & $0.286\pm0.063$ & \dots  & \dots & \dots\\
A2744-ID11\tablefootmark{b} & 00:14:22.63 & $-$30:23:30.45 & 4.5 & $0.253\pm0.056$ & \dots  & \dots & \dots\\
MACSJ0416-ID01\tablefootmark{a} & 04:16:10.79 & $-$24:04:47.53 & 15.4 & $1.010\pm0.066$ & $1.319\pm0.103$ & $0.23\pm0.06$ & 2.086\tablefootmark{c}\\
MACSJ0416-ID02\tablefootmark{a} & 04:16:06.96 & $-$24:03:59.96 & 6.8 & $0.406\pm0.062$ & $0.574\pm0.132$ & $0.32\pm0.15$ & 1.953\tablefootmark{c}\\
MACSJ0416-ID03\tablefootmark{a} & 04:16:08.81 & $-$24:05:22.58 & 5.8 & $0.389\pm0.067$ & $0.411\pm0.072$ & $0.10\pm0.07$ & $1.29_{-0.39}^{+0.11}$\tablefootmark{e}\\
MACSJ0416-ID04\tablefootmark{a} & 04:16:11.67 & $-$24:04:19.44 & 5.1 & $0.333\pm0.066$ & $0.478\pm0.166$ & $0.37\pm0.21$ & $2.27_{-0.61}^{+0.17}$\tablefootmark{e}\\
MACSJ0416-ID05\tablefootmark{b} & 04:16:10.52 & $-$24:05:04.77 & 4.6 & $0.302\pm0.066$ & \dots & \dots & 1.849\tablefootmark{f}\\
MACSJ1149-ID01\tablefootmark{a} & 11:49:36.09 & +22:24:24.60 & 5.9 & $0.442\pm0.074$ & $0.579\pm0.134$ & $0.28\pm0.13$ & 1.46\tablefootmark{c}\\
MACSJ1149-ID02\tablefootmark{b} & 11:49:40.32 & +22:24:42.00 & 4.6 & $0.524\pm0.113$ & \dots & \dots & \dots\\
MACSJ1149-ID03\tablefootmark{b} & 11:49:35.41 & +22:23:38.60 & 4.5 & $0.326\pm0.072$ & \dots & \dots & \dots\\
A370-ID01\tablefootmark{a} & 02:39:56.57 & $-$01:34:26.40 & 25.1 & $1.624\pm0.065$ & $1.925\pm0.088$ & $0.22\pm0.03$ & 1.06\tablefootmark{f}\\
A370-ID02\tablefootmark{a} & 02:39:57.58 & $-$01:34:53.67 & 9.1 & $1.656\pm0.182$ & $1.778\pm0.192$ & $0.10\pm0.05$ & $2.581_{-0.180}^{+0.193}$\tablefootmark{g}\\
A370-ID03\tablefootmark{a} & 02:39:58.15 & $-$01:34:24.52 & 7.0 & $0.655\pm0.094$ & $1.007\pm0.205$ & $0.42\pm0.13$ & $1.258$\tablefootmark{h}\\
A370-ID04\tablefootmark{b} & 02:39:53.25 & $-$01:33:29.26 & 4.7 & $0.439\pm0.094$ & \dots & \dots & \dots\\
A370-ID05\tablefootmark{b} & 02:39:53.20 & $-$01:34:38.16 & 4.5 & $0.283\pm0.063$ & \dots & \dots & \dots\\
AS1063-ID01\tablefootmark{a} & 22:48:49.06 & $-$44:32:25.39 & 89.3 & $6.295\pm0.071$ & $6.806\pm0.119$ & $0.14\pm0.01$ & $2.511_{-0.113}^{+0.111}$\tablefootmark{g}\\
AS1063-ID02\tablefootmark{a} & 22:48:47.29 & $-$44:30:48.06 & 10.9 & $1.282\pm0.118$ & $1.273\pm0.142$ & $0.10\pm0.05$ & 1.438\tablefootmark{f}\\
AS1063-ID03\tablefootmark{a} & 22:48:46.58 & $-$44:30:46.83 & 5.2 & $0.648\pm0.124$ & $1.396\pm0.399$ & $0.45\pm0.14$ & $1.468_{-0.063}^{+0.065}$\tablefootmark{g}\\
AS1063-ID04\tablefootmark{b} & 22:48:41.79 & $-$44:31:57.80 & 4.9 & $0.338\pm0.069$ & \dots & \dots & 0.611\tablefootmark{i}\\
AS1063-ID05\tablefootmark{b} & 22:48:42.24 & $-$44:32:29.20 & 4.6 & $0.320\pm0.069$ & \dots & \dots & \dots\\
\hline
\end{tabular}
\tablefoot{Column 1: Source ID. Columns 2, 3: Centroid J2000 position of ID. Column 4: Signal-to-noise of the detection, i.e., the ratio of peak intensity and background rms. Column 5: PB-corrected peak intensity and $1\sigma$ error. Column 6: PB-corrected integrated flux density and $1\sigma$ error from $uv$ fitting. Column 7: Effective radius and $1\sigma$ error from $uv$ fitting, defined as $r_s=\sqrt{a_s\times b_s}$ (half-light radius). Column 8: Source redshift.
\tablefoottext{a}{High-significance ($\textmd{S/N}\geq5$) detections.}
\tablefoottext{b}{Low-significance ($4.5\leq\textmd{S/N}<5$) detections. Instead of performing a $uv$ fitting, we estimate the integrated flux density using the peak intensity and assuming given source size parameters (see $\S$\ref{sect_obs}). For sources that lack redshift estimates, we assume a Gaussian redshift distribution with mean $z=2$ and $\sigma=0.5$.}
\tablefoottext{c}{Spectroscopic redshift from GLASS, already noted in Papers II and IV}.
\tablefoottext{d}{Spectroscopic redshift from Bauer et al. (in prep.).}
\tablefoottext{e}{Photometric redshift found in Paper II.}
\tablefoottext{f}{Spectroscopic redshift from GLASS.}
\tablefoottext{g}{Photometric redshift from HFF-DeepSpace.}
\tablefoottext{h}{Spectroscopic redshift from \citet{Lagattuta2022}.}
\tablefoottext{i}{Spectroscopic redshift from CLASH.}
Photometric redshifts best value and $1\sigma$ error from SED fitting are presented here only for reference, as we use the full probability distribution found for each photometric redshift.}
\label{tab_snr45gals}
\end{center}
\end{table*}

Five out of the six FF galaxy clusters were observed by ALMA in Band 6, as part of the ALMA Frontier Fields Survey (programs \#2013.1.00999.S in cycle 2 and \#2015.1.01425.S in cycle 3, PI: F. Bauer). These clusters are Abell 2744, MACSJ0416.1-2403, MACSJ1149.5+2223, Abell 370 and Abell S1063 (hereafter A2744, MACSJ0416, MACSJ1149, A370 and AS1063 respectively). Paper I introduces the $1.1\,\textmd{mm}$ mosaic images, data reduction and analysis of the cycle 2 observations, which cover the first three galaxy clusters. Paper III presents a summary of cycle 3 observations, which cover A370 and AS1063. For the five clusters, the inner $\approx2'\times2'$ FF regions were covered with 126 pointings each, with a Local Oscillator frequency set to $263.14\,\textmd{GHz}$ ($\approx1.1\,\textmd{mm}$). Datasets were reduced using the Common Astronomy Software Applications (CASA; \citealt{McMullin2007}, and imaged using the multi-frequency synthesis algorithm in CLEAN, adopting a $0\farcs1$ pixel size. Cleaned images were made using niter = 1000 and a threshold of four times the rms measured in the dirty images (which were generated using niter = 0), applying no masking during cleaning. Mosaics were created using natural weighting (see details in Paper I). Table \ref{tab_clus} lists the cluster properties and natural-weighted mosaic details for the five clusters (images are shown in Fig. \ref{fig_maps}). Since the sixth FF cluster, MACSJ0717.5+3745, was only partially observed in cycle 3 to much lower sensitivity, source extraction was not performed on it, and thus we do not include it in this study.

We detect sources by searching for pixels with signal-to-noise ratio $(\textmd{S/N})$ larger or equal than a given threshold, grouping pixels as individual sources with the DBSCAN python algorithm \citep{Pedregosa2012}. We choose a threshold of $\textmd{S/N}=4.5$, as this is found to be an appropriate tradeoff between the purity and completeness correction factors that will be applied to our number counts (see $\S$\ref{sect_method}). Detections in the first three FF clusters were already reported in Paper I and Paper IV, making a total of 19 sources. Here we introduce our detections in A370 and AS1063. In each mosaic, we perform the source extraction within the region where the primary beam (PB) sensitivity is at least a given fraction of the peak sensitivity. In Paper IV, we considered $\textmd{PB}>0.5$. However, in one of the last two FFs we find a high $\textmd{S/N}$ source at $\textmd{PB}=0.34$, A370-ID02. Therefore, in this study we use $\textmd{PB}>0.3$ for all cluster fields. This does not add sources to the first three fields. Throughout this paper, all source flux densities and peak intensities correspond to PB-corrected values.

Extracting sources at $\textmd{S/N}\geq4.5$ in the last two FFs adds ten new detections to our catalog (see Table \ref{tab_snr45gals}). In each of these cluster fields, we detect three high-significance ($\textmd{S/N}\geq5$) and two low-significance ($4.5\leq\textmd{S/N}<5$) sources. Detection coordinates are shown in Fig. \ref{fig_maps}. Detections have peak intensities in the range $\sim0.28-6.30\,\textmd{mJy}\,\textmd{beam}^{-1}$. We measure integrated flux densities in the \textit{uv}-plane for our new high-significance detections as in Paper I, using the UVMCMCFIT python algorithm \citep{Bussmann2016}. These Gaussian fits give integrated flux densities in the range $\sim1.00-6.81\,\textmd{mJy}$, effective radii in the range $\lesssim0\farcs10-0\farcs45$ (half-light radii\footnote{The half-light size of a source is given by 2.35 times its effective radius.}) and axial ratios in the range $\sim0.36-0.80$.

As in Paper IV, we estimate the integrated flux densities of low-significance detections in A370 and AS1063 using source peak intensities; this choice is motivated by the large uncertainties in integrated source flux densities that are given at low S/N by two-dimensional Gaussian fits in the $uv$-plane. We adopt as their observed effective radius and axial ratio the median values found for the high-significance sources. Including the five FFs, these are $r_{\textmd{eff,obs}}=0\farcs23$ and $q_{\textmd{obs}}=0.58$, respectively. We conduct source injection simulations (see $\S$\ref{sect_method}), finding a typical ratio between the peak and integrated flux density for these size parameters of 0.80 and 0.71 in A370 and AS1063, respectively. These factors give image-plane integrated flux densities of the $4.5\leq\textmd{S/N}<5$ detections ranging from $\sim0.35$ to $\sim0.55\,\textmd{mJy}$.

Each cluster field covers an observed area of $\sim5.2\,\textmd{arcmin}^2$ ($\textmd{PB}>0.3$ region). The five clusters sum to a total image-plane area of $\sim26\,\textmd{arcmin}^2$. This corresponds to $\sim6$ times the area of the Hubble Ultra Deep Field (HUDF, \citealt{Dunlop2017}), $\sim26$ times the initial ALMA Spectroscopic Survey in the HUDF (ASPECS-Pilot, \citealt{Walter2016,Aravena2016}) and $\sim6$ times the ASPECS-LP \citep{Aravena2020}. Note, however, that this area is reduced when lens models are applied (see $\S$\ref{sect_models}).

\subsection{Source redshifts}\label{sect_z}

For our ALMA detections in the first three FFs, we start by adopting the same redshifts as in Paper IV. These include spectroscopic redshifts from the GLASS survey using grism spectroscopy at HST \citep{Treu2015}, and photometric redshifts (including full probability distributions) derived in Paper II for our high-significance detections via spectral energy distribution (SED) fitting. Since the publication of Paper IV, various new catalogs have been released, providing recent estimates for these galaxy clusters. These catalogs include: photometric redshifts reported by \citet{Ishigaki2018}, the HFF-DeepSpace project \citep{Shipley2018}, \citet{Bhatawdekar2019}, the KLASS survey \citep{Mason2019}, the SHARDS FF survey \citep{Griffiths2021}, \citet{Pagul2021}, the ALCS survey \citep{Kokorev2022}, and the UNCOVER survey \citep{Weaver2023}; and spectroscopic redshifts by \citet{DeLaVieuville2019}, \citet{Vanzella2021}, and \citet{Richard2021}. We search for counterparts in these catalogs to those of our sources that lack spectroscopic or photometric estimates. We check that none of these sources have reliable counterparts within $\approx1''$ from their peak positions.

Bauer et al. (in prep.) performed an ALMA linescan targeting our high-significance ALMA detections in A2744 (program \#2017.1.01219.S in cycle 5, PI: F. Bauer), obtaining several spectroscopic redshift estimates. We then replace photometric redshifts for A2744 detections with secure spectroscopic redshifts when available. New included redshifts and secure line(s) identifications are as follows A2744-ID02 has $z_{\textmd{spec}}=2.585$ from CO (3-2) and CO (5-4) lines (previously in Paper IV we adopted $z_{\textmd{spec}}=2.482$ from GLASS, which was based on a red continuum); A2744-ID03 has $z_{\textmd{spec}}=3.058$ from CO (4-3) and CO (5-4) lines;
A2744-ID04 has $z_{\textmd{spec}}=1.498$ from CO (2-1), CO (4-3) and [C I] (1-0) lines;
A2744-ID05 has $z_{\textmd{spec}}=2.409$ from CO (3-2), [C I] (1-0), CO (7-6) and [C I] (2-1) lines; and
A2744-ID07 has $z_{\textmd{spec}}=2.644$ from CO (3-2), CO (5-4) and CO (8-7) lines.

We search for counterparts to our ALMA detections in the last two FFs considering several public catalogs that report photometric and/or spectroscopic redshift estimates. These include: photometric redshifts estimated by \citet{Monna2014}, the CLASH team \citep{Postman2012,Molino2017}, \citet{Ishigaki2018}, the HFF-DeepSpace project, the KLASS survey, the ASTRODEEP survey \citep{Bradac2019}, the SHARDS FF survey, \citet{Pagul2021}, and the ALCS survey; catalogs of spectroscopic redshifts by \citet{Richard2010}, \citet{Gomez2012}, \citet{Boone2013}, \citet{Johnson2014}, \citet{Richard2014}, \citet{Diego2016}, \citet{Lagattuta2017}, \citet{Vanzella2017}, \citet{Diego2018}, \citet{Strait2018}, \citet{Lagattuta2019}, \citet{Vega-Ferrero2019}, \citet{Walth2019}, \citet{Richard2021}, \citet{Lagattuta2022}, the GLASS survey, and the CLASH survey using VIMOS \citep{Balestra2013} and MUSE \citep{Karman2015,Caminha2016,Karman2017,Mercurio2021} at VLT; and redshift estimates for \textit{Herschel} detections \citep{Rawle2016}.

Two sources in A370 and two in AS1063 have spectroscopic redshifts, these are A370-ID01, A370-ID03, AS1063-ID02 and AS1063-ID04. A370-ID01 has a $z_{\textmd{spec}}=1.06$ (secure) counterpart within $\approx0\farcs4$ from GLASS, identifying multiple emission lines. A370-ID03 has a $z_{\textmd{spec}}=1.258$ (secure) counterpart within $\approx0\farcs2$ from \citet{Lagattuta2022} using MUSE at VLT. They give a redshift confidence flag of 3 to this source, indicating high reliability; this means that the redshift is either based on multiple clear spectral features, or on a single high-significance emission line with additional information. AS1063-ID02 has a $z_{\textmd{spec}}=1.438$ (possible/probable) counterpart within $\approx0\farcs2$ from GLASS. \citet{Treu2015} give a redshift quality flag of 2.5 to this source, indicating that either a single strong emission line is robustly detected and the identification is supported by a photometric redshift, or that more than one feature is marginally detected, or that a single line is detected with marginal quality. AS1063-ID04 has a $z_{\textmd{spec}}=0.61$ counterpart within $\approx1\farcs1$ from \citet{Gomez2012}, obtained from $\textmd{H}\beta$ and $\textmd{[O III]}$ emission lines using GMOS at Gemini-South. It has a $z_{\textmd{spec}}=0.609$ (possible) counterpart within $\approx0\farcs9$ from GLASS, having a single line detection of marginal quality but strong $\textmd{[S III]}\lambda9069$ and $\textmd{[S III]}\lambda9531$ emission lines using the G141 grism. It also has a $z_{\textmd{spec}}=0.611$ (secure) counterpart within $\approx0\farcs9$ from CLASH using MUSE at VLT. We adopt the latter value for this work. This counterpart is well resolved in HST images, displaying a clear spiral-like morphology. Using CLASH HST mosaics, \citet{Connor2017} derived a semi-major axis length of $\approx3\farcs5$ for this galaxy. AS1063-ID04 is found towards the disk of this galaxy, at $\approx1''$ from its center. In addition, this counterpart was listed in a follow-up study to $24\,\mu\textmd{m}$ selected galaxies by \citet{Walth2019}. They identified eight emission lines using LDSS-3 at Magellan-Clay in this galaxy, and reported the discovery of a luminous kpc-sized HII region at one edge of its disk.

We restrict our photometric redshift search to those data providing full posterior probability distributions, i.e. the HFF-DeepSpace project. \citet{Shipley2018} derived photometric redshifts using up to 22 filters in the wavelength range $0.2-8\,\mu\textmd{m}$ and the EAZY fitting code \citep{Brammer2008}. Among the sources that lack spectroscopic redshifts, one in A370 and two in AS1063 have counterparts with reliable photometric redshifts within $0\farcs5$. We use the redshift quality parameter $p_{\Delta z}$ provided by the catalog as a measure of the $z_{\textmd{phot}}$ reliability, with $p_{\Delta z}\approx1$ indicating that most of the total integrated probability lies within $\Delta z=0.2$ of the $z_{\textmd{phot}}$ estimate. Listing the best (i.e. $z_{\textmd{m}2}$) redshift values and $1\sigma$ errors of the closest counterpart as a reference, these sources are A370-ID02 ($z_{\textmd{phot}}=2.581_{-0.180}^{+0.193}$), AS1063-ID01 ($z_{\textmd{phot}}=2.511_{-0.113}^{+0.111}$) and AS1063-ID03 ($z_{\textmd{phot}}=1.468_{-0.063}^{+0.065}$). All these estimates have a redshift quality parameter $p_{\Delta z}\approx1$. AS1063-ID05 has a counterpart within $\approx0\farcs3$ having $z_{\textmd{phot}}=2.256_{-1.302}^{+1.766}$. However, it has $p_{\Delta z}\approx0.2$, indicating a broad and/or multi-modal probability distribution. Therefore, we do not include this redshift estimate in this work.

We note that one detection in A370 (A370-ID03) and three in AS1063 (AS1063-ID01, AS1063-ID02 and AS1063-ID04) have counterparts in the \textit{Herschel} catalog within $\approx1''$ (source coordinates given by either VLA or HST position references). A370-ID03 has a IR-estimated redshift of $\sim0.9$ from SED fitting using five \textit{Herschel} bands (see \citealt{Rawle2016}); we do not use this estimate, as we prioritize the spectroscopic redshift found for A370-ID03. We also search for counterparts to our sources in available catalogs of FF cluster members, as well as in catalogs used as input by the lens models included in this work. We find that none of our ALMA detections have counterparts in them.

We list the source redshifts chosen for this work in Table \ref{tab_snr45gals}. When available, we use spectroscopic redshifts and (if not) photometric redshift probability distributions. Nine detections have no reliable counterparts in the aforementioned catalogs. As in Paper IV, for these sources we assume a Gaussian redshift distribution centered at $z=2$ with $\sigma=0.5$, which is consistent within $1\sigma$ with average redshift values found for faint dusty galaxies from the literature (e.g. \citealt{Aravena2016,Dunlop2017,Aravena2020}).

\subsection{Lensing models}\label{sect_models}

\begin{table*}
\begin{center}
\caption{Lensing models considered in this work.}
\begin{tabular}{cc}
\hline \hline
Model & References\\ \hline
Bradac v4\tablefootmark{a}, v4.1\tablefootmark{a} & \citet{Bradac2005,Bradac2009,Strait2018}\\
Caminha v4\tablefootmark{b} & \citet{Caminha2017}\\
CATS v4, v4.1\tablefootmark{c} & \citet{Jullo2009,Richard2014,Jauzac2014,Jauzac2015b,Jauzac2016};\\
& \citet{Lagattuta2017,Lagattuta2019}\\
Diego v4, v4.1 & \citet{Diego2005,Diego2007,Diego2015,Diego2016,Diego2018,Vega-Ferrero2019}\\
GLAFIC v4\tablefootmark{d} & \citet{Oguri2010,Kawamata2016,Kawamata2018}\\
Keeton v4 & \citet{Keeton2010,Ammons2014,McCully2014}\\
Sharon v4 & \citet{Jullo2007,Johnson2014}\\
Williams v4, v4.1\tablefootmark{e} & \citet{Liesenborgs2006,Liesenborgs2007,Sebesta2016}\\
\hline
\end{tabular}
\tablefoot{All models cover the region where our ALMA sources lie, except for GLAFIC v4 in A370 (thus not included for this cluster).
\tablefoottext{a}{Only available for A370.}
\tablefoottext{b}{Only available for MACSJ0416.}
\tablefoottext{c}{Not available for A370.}
\tablefoottext{d}{Not available for MACSJ1149.}
\tablefoottext{e}{Only available for A370 and AS1063.}
}
\label{tab_models}
\end{center}
\end{table*}

We correct our number counts by gravitational lensing effects making use of the models publicly available in the FF website\footnote{\url{http://archive.stsci.edu/prepds/frontier/lensmodels/}}, which are produced by independent teams. These models provide maps of the normalized mass surface density and shear of the galaxy cluster for a redshift $z=\infty$ background. We use the full set of individual mass reconstructions released by the teams (which sample  the range of uncertainties), reprojected to the size and resolution of the ALMA maps. From these maps and adopting a given source redshift, we can compute magnification maps, deflection fields around each galaxy cluster and thus effective source-plane areas for each detection (i.e., the angular area where a map pixel with a given peak intensity can be detected over a S/N threshold). We refer the reader to Paper IV for more details regarding how these quantities are obtained from the aforementioned maps.

For the first three FFs, we use the same lens models as in Paper IV, as no newer versions have been publicly released\footnote{We note that the FF website released corrected Sharon v4 models after acceptance of Paper IV. In this work (as well as in the Paper IV Corrigendum) we use updated models.}. Table \ref{tab_models} lists these models, together with the models considered for the two FF clusters introduced in this work; we adopt for use eight, nine, seven, nine, and nine lens models for A2744, MACSJ0416, MACSJ1149, A370, and AS1063, respectively. We consider only v4 or newer models, as these use the best data to date as constraints\footnote{On top of those listed, Zitrin \& Merten (e.g., \citealt{Zitrin2015}) also supplied lens models for the FF clusters, but their latest update was v3 so these were not used here.}. We also restrict our set only to models that cover the whole region where our ALMA sources lie. A370-ID03 is not covered by GLAFIC v4 model, and therefore we do not include it for A370. A fraction of the region where the ALMA maps have $\textmd{PB}>0.3$ is not fully covered by the models GLAFIC v4 ($\sim2\%$ and $3\%$ for A2744 and AS1063, respectively), Sharon v4 ($\sim0.7\%$ for A370), Williams v4 ($\sim3\%$, $19\%$, $10\%$ and $14\%$ for A2744, MACSJ0416, MACSJ1149 and A370, respectively) and Williams v4.1 ($\sim14\%$ for A370). We consider no magnification effects ($\mu=1$) for the missing pixels in magnification maps.

\section{Methodology}\label{sect_method}

We compute demagnified number counts at $1.1\,\textmd{mm}$ considering ALMA detections down to $\textmd{S/N}=4.5$. We follow the procedure that is described in detail in Paper IV. There, we use a Monte Carlo approach to take into account the uncertainties in observed flux densities, adopted redshifts, and magnifications given by the lens models. After running simulation realizations for all lens models, we compute median values as the best estimates of the true values for our number counts.

We conduct this procedure for the five galaxy clusters with no modifications other than the PB limit, restricting to the $\textmd{PB}>0.3$ region instead of using $\textmd{PB}>0.5$ (see $\S$\ref{sect_obs}). For the first three FFs, this implies recomputing the quantities where the PB limit is involved. These are: i) completeness, that is, the proportion of sources that were not detected because their noise level shifted their peak S/N below our chosen threshold; ii) flux deboosting \citep{Hogg1998,Weiss2009}; iii) fraction of spurious sources, i.e., generated by noise; and iv) effective area as a function of demagnified peak intensity. The first three of these corrections are done in the image plane.

\begin{figure}
\centering
\resizebox{0.75\hsize}{!}{\includegraphics{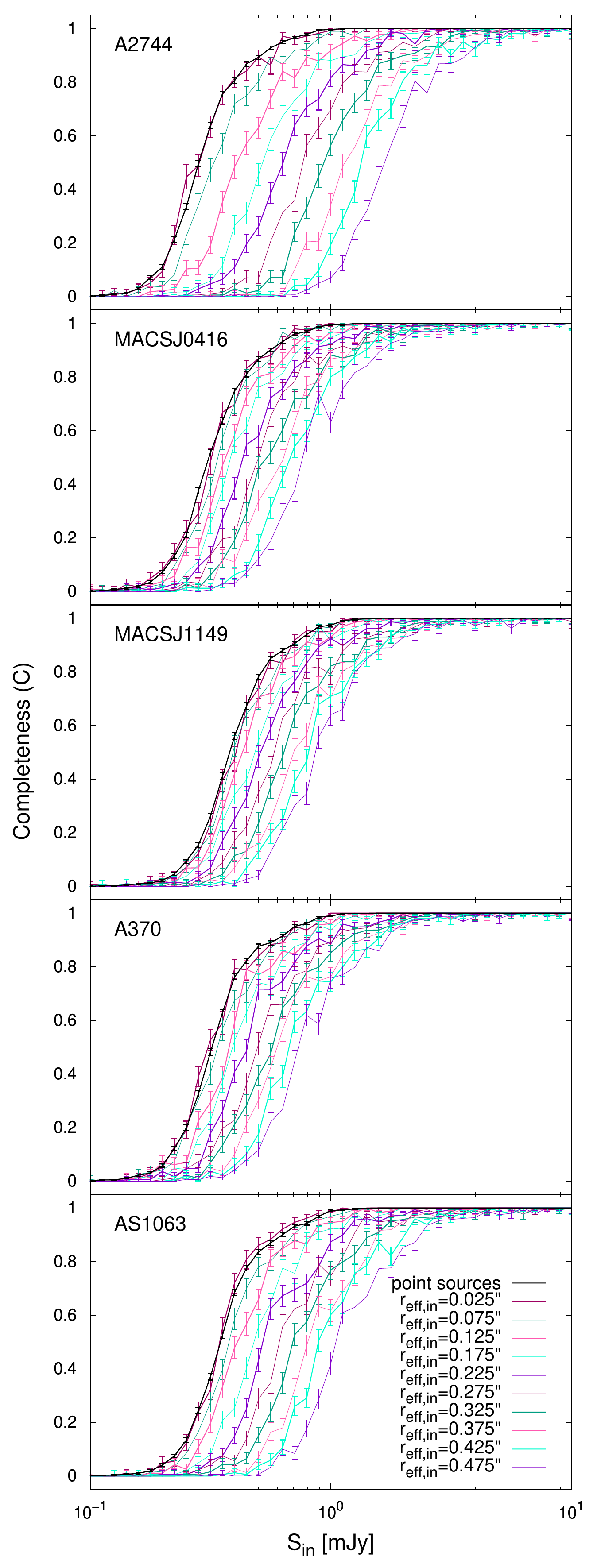}}
\caption{Completeness correction $C$ as a function of image-plane integrated flux density and separated in bins of image-plane scale radius (half-light radius). Error bars indicate binomial confidence intervals.}
\label{fig_comp_fluxout_reffin}
\end{figure}

\begin{figure}
\centering
\resizebox{0.75\hsize}{!}{\includegraphics{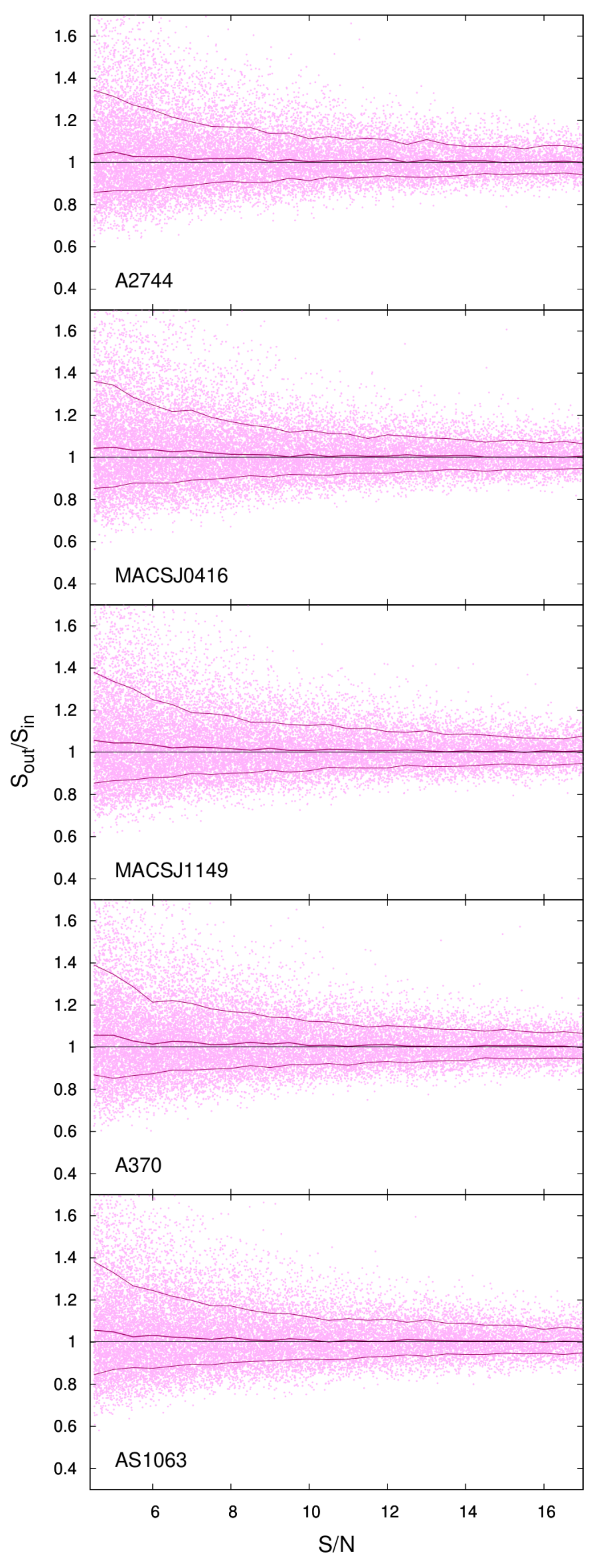}}
\caption{Deboosting correction as a function of S/N. We display the ratio between the extracted and injected flux densities for our simulated sources as light magenta dots. Thick dark magenta lines correspond to median values while thin dark magenta lines indicate the 16th and 84th percentiles.}
\label{fig_debcorr}
\end{figure}

We use source injection simulations, in the image-plane, to estimate the completeness and deboosting corrections (see Paper IV $\S$3.1 for details). After source extraction, we compute the completeness as a function of image-plane integrated flux density $S_{\textmd{obs}}$ and separate it in bins of image-plane scale radius. Completeness corrections for the five cluster fields are shown in Fig. \ref{fig_comp_fluxout_reffin}. For point sources, a value of $50\%$ is reached at image-plane flux densities of $0.32\,\textmd{mJy}$ for A370 and $0.35\,\textmd{mJy}$ for AS1063. However, the completeness drops to $19\%$ and $7\%$ respectively at the same flux densities for image-plane source sizes in the range $0\farcs20-0\farcs25$ (i.e., for the image-plane size assumed for our low-significance detections). Figure \ref{fig_debcorr} shows the ratio between the extracted and injected flux densities for our simulations. We find that at $\textmd{S/N}=4.5$ the noise boosts the flux densities by $6\%$ for both A370 and AS1063. We correct for this effect both the observed peak intensities and integrated flux densities of all our detections.

We obtain a rough estimate of the Eddington bias (i.e., an overestimate of the derived faint-end number counts due to noise fluctuations and a steep underlying flux density distribution; \citealt{Eddington1913}) as done in Paper IV, making no assumptions regarding the true underlying distribution of flux densities, since the number density of ALMA sources in the FFs is low even after adding two clusters. We create sets of $10^4$ simulated point sources that follow the SIDES galaxy formation model \citep{Bethermin2017} distributions of both redshift and $1.1\,\textmd{mm}$ flux density, lens these sources using the ``best'' CATS v4 model for each cluster, inject and extract them in our ALMA mosaics down to $\textmd{S/N}=4.5$, and obtain the ratio between output and input demagnified flux density as a function of S/N (see Paper IV $\S$3.1 for details). At $\textmd{S/N}=4.5$, we find flux enhancements by 21\%, 29\%, 22\%, 19\% and 33\% for A2744, MACSJ0416, MACSJ1149, A370, and AS1063, respectively. These ratios are consistent with the deboosting corrections obtained in Fig. \ref{fig_debcorr}. We thus consider safe to skip Eddington bias corrections in this work.

\begin{figure}
\centering
\resizebox{0.8\hsize}{!}{\includegraphics{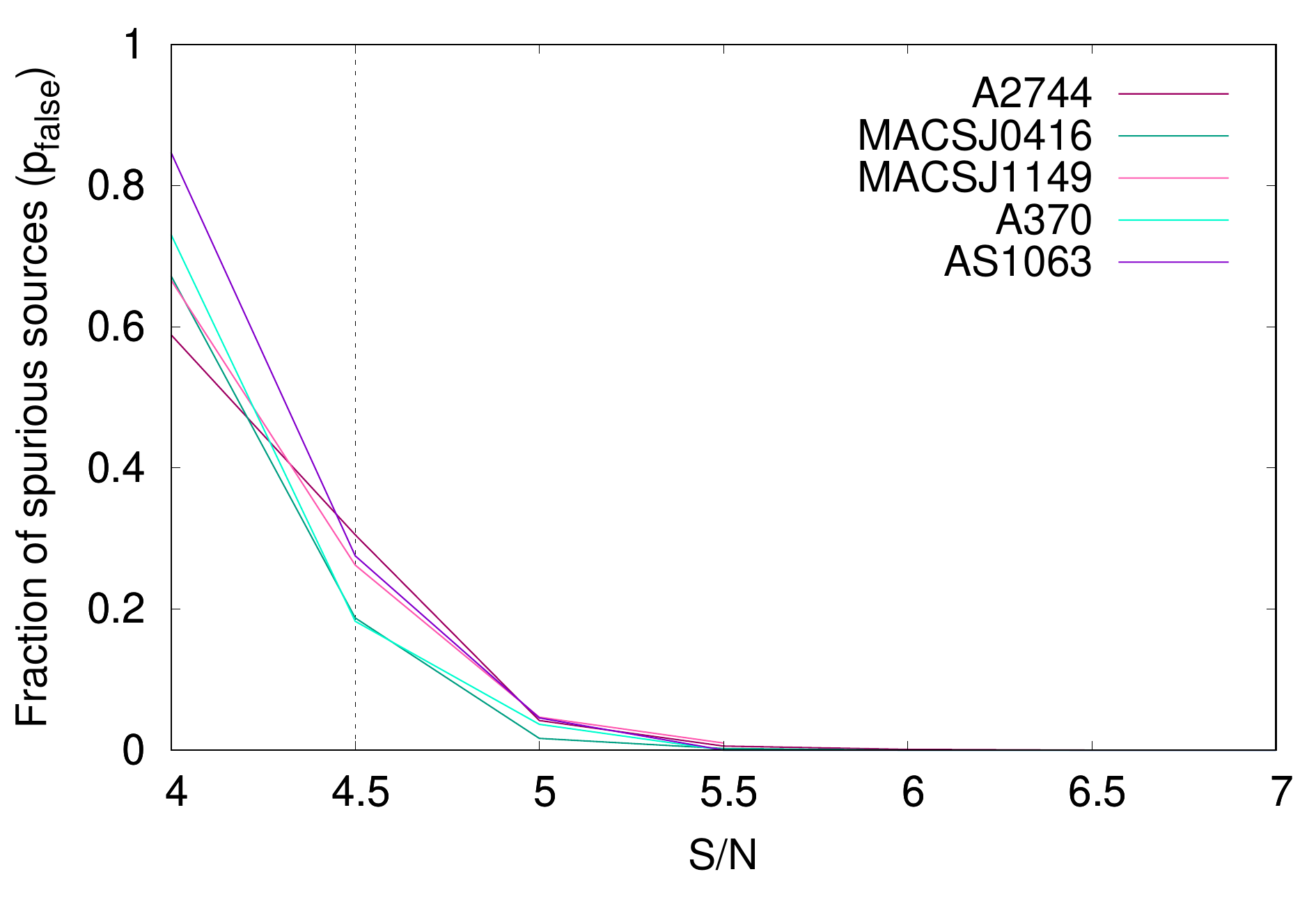}}
\caption{Fraction of spurious sources at a given S/N. We display curves for A2744, MACSJ0416, MACSJ1149, A370 and AS1063 in red, green, blue, magenta, and cyan, respectively. A vertical dotted line indicates our S/N threshold of 4.5.}
\label{fig_fsp}
\end{figure}

We compute the fraction of spurious sources as a function of S/N for each galaxy cluster, $p_{\textmd{false}}$, as the average ratio between the number of detections over a peak S/N in a set of 300 simulated noise maps and in the actual ALMA mosaic (see Paper IV and Paper IV Corrigendum). Figure \ref{fig_fsp} shows this fraction for each of the five FFs. At $\textmd{S/N}\geq4.5$ $p_{\textmd{false}}$ is $\approx18\%$ for A370 and $\approx28\%$ for AS1063. At $\textmd{S/N}\geq5$, it drops to $\approx4\%$ and $\approx5\%$ respectively. Based on the source extraction on the simulated noise maps, the average number of spurious sources at $\textmd{S/N}\geq4.5$ is $0.93\pm1.33$ for A370 and $1.45\pm1.79$ for AS1063. This is consistent within $1\sigma$ with both the amount of spurious sources from the negative mosaics (zero and one, respectively) and the number of sources without reliable counterparts at other wavelengths (two and one, respectively).

For a given galaxy cluster, source redshift and individual mass reconstruction of a lens model, the effective source-plane area over which a source can be detected is obtained as follows. We use the surface mass density and shear maps to obtain a magnification map. Both this and the PB-corrected rms map for the cluster are deflected to the source plane. At each demagnified peak intensity $S_{\textmd{demag,peak}}$, the effective area $A_{\textmd{eff}}$ is obtained from the source-plane pixels that satisfy $S_{\textmd{demag,peak}}/\sigma_{\textmd{demag}}\geq4.5$, with $\sigma_{\textmd{demag}}$ the PB-corrected rms corrected for magnification. Since we need to create several source-plane maps for each lens model (sampling the source redshift distributions) but also want our Monte Carlo simulations to run fast, we precompute source-plane maps for all mass reconstructions and lens models, over the redshift range $z_{\textmd{min}}=0.4$ to $z_{\textmd{max}}=4$, using steps of $\Delta z=0.2$. We use these maps to obtain curves of $A_{\textmd{eff}}$ as a function of $S_{\textmd{demag,peak}}$ for the five FFs.

We run 1000 realizations of the number counts per lens model per cluster field. Each is computed from a simulated source catalog, using the same centroid coordinates and observed source sizes as the 29 ALMA detections. Simulated catalogs have, for each source $i$, integrated flux densities $S_{\textmd{obs},i}$ and peak intensities $S_{\textmd{obs,peak},i}$ drawn from Gaussian distributions centered in the measured values (see Table \ref{tab_snr45gals}), and redshifts $z_i$ drawn from the values/distributions selected in $\S$\ref{sect_z}. From the simulated $S_{\textmd{obs},i}$ and $S_{\textmd{obs,peak},i}$ values, we obtain completeness $C_i$ (for each source size) and deboosting corrections, $(\textmd{S/N})_i$ values (later considering only $(\textmd{S/N})_i\geq4.5$ sources) and fractions of spurious sources $p_{\textmd{false},i}$. This is done interpolating the curves shown in Figs. \ref{fig_comp_fluxout_reffin} to \ref{fig_fsp}.

From the source centroid pixel coordinates and $z_i$, and selecting randomly one of the mass reconstructions provided by the lens model, we obtain each source magnification $\mu_i$. We neglect the effects of differential magnification, as all our detections lie far enough (i.e. more than four synthesized beams away) from critical lines in most of the adopted redshifts and lens models\footnote{However, see Paper IV for a discussion about A2744-ID09 and A2744-ID11.}. With all these quantities, we obtain demagnified integrated flux densities $S_{\textmd{demag},i}=S_{\textmd{obs},i}/\mu_i$, demagnified peak intensities $S_{\textmd{demag,peak},i}=S_{\textmd{obs,peak},i}/\mu_i$, and source effective areas $A_{\textmd{eff},i}$. The latter is computed interpolating the $A_{\textmd{eff}}$ vs. $S_{\textmd{demag,peak}}$ curves that correspond to the same mass reconstruction used to obtain $\mu_i$.

We then compute differential (cumulative) number counts at each demagnified flux density bin (limit) as
\begin{equation}
\frac{\textmd{d}N}{\textmd{d}\log(S)}=\frac{1}{\Delta\log(S)}\sum\limits_{i}X_i, \label{eq_ndiff}
\end{equation}
\noindent and
\begin{equation}
N(>S)=\sum\limits_{i}X_i, \label{eq_ncumu}
\end{equation}
\noindent where we sum the contribution $X_i$ to the counts by the sources having $S_{\textmd{demag},i}$ within (larger than) that flux density bin (limit), and
\begin{equation}
X_i=\frac{1-p_{\textmd{false},i}}{C_i\,A_{\textmd{eff},i}}. \label{eq_xis}
\end{equation}

We finally obtain combined differential (cumulative) counts, both in each separate cluster and including the five FFs, as the median values per flux density bin (limit). The associated uncertainties are given by the 16th and 84th percentiles, added in quadrature with scaled Poisson confidence levels for $1\sigma$ lower and upper limits \citep{Gehrels1986}.

\begin{figure*}
\centering
\resizebox{0.95\hsize}{!}
{\includegraphics{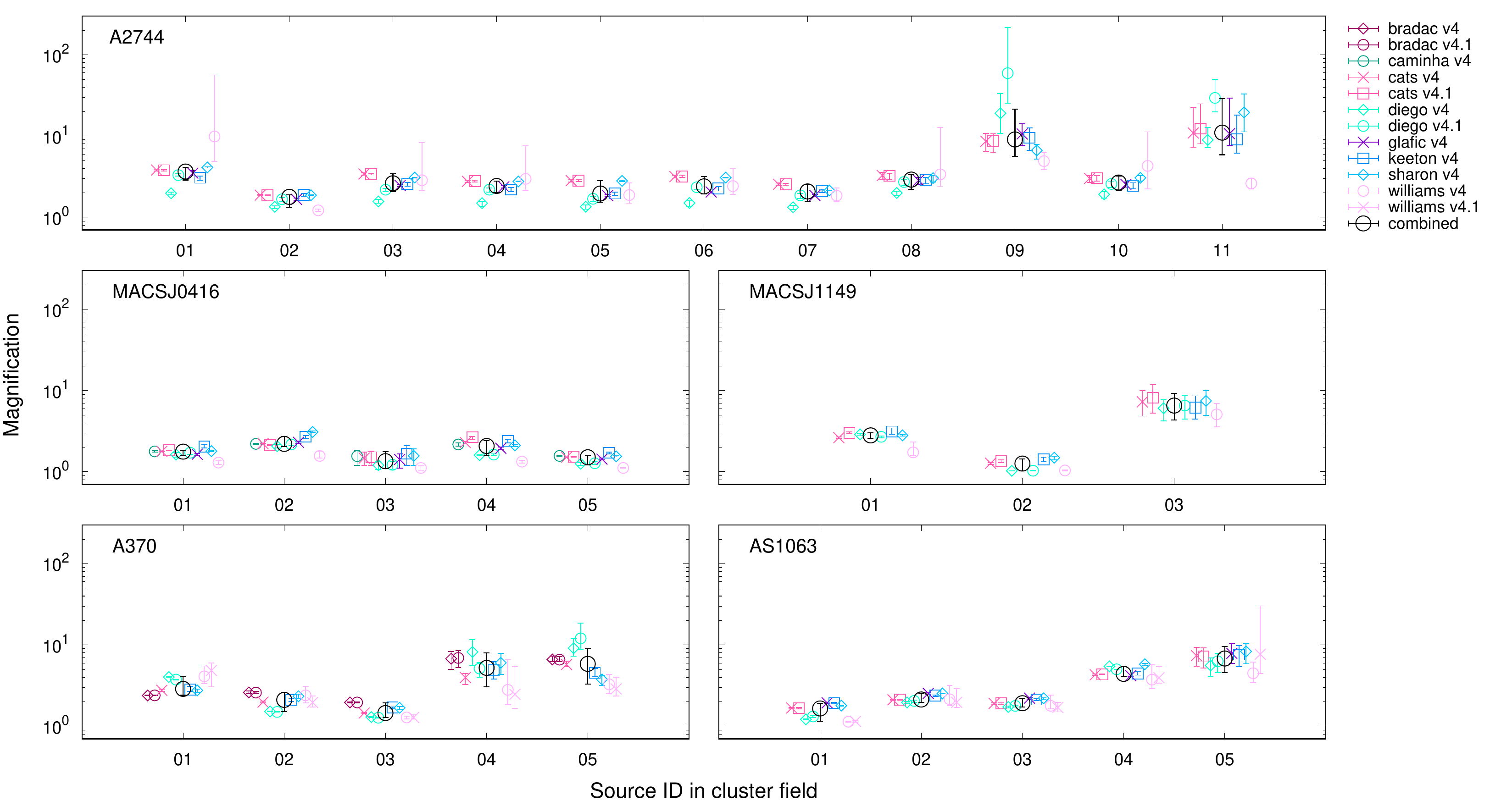}}
\caption{Median magnification per source for the lens models (colored symbols), and also combining all models for each cluster field (large black circles). Error bars indicate the 16th and 84th percentiles. Values for each model have been offset around the source ID for clarity.}
\label{fig_magnif_gal}
\end{figure*}

\begin{figure}
\centering
\resizebox{0.8\hsize}{!}
{\includegraphics{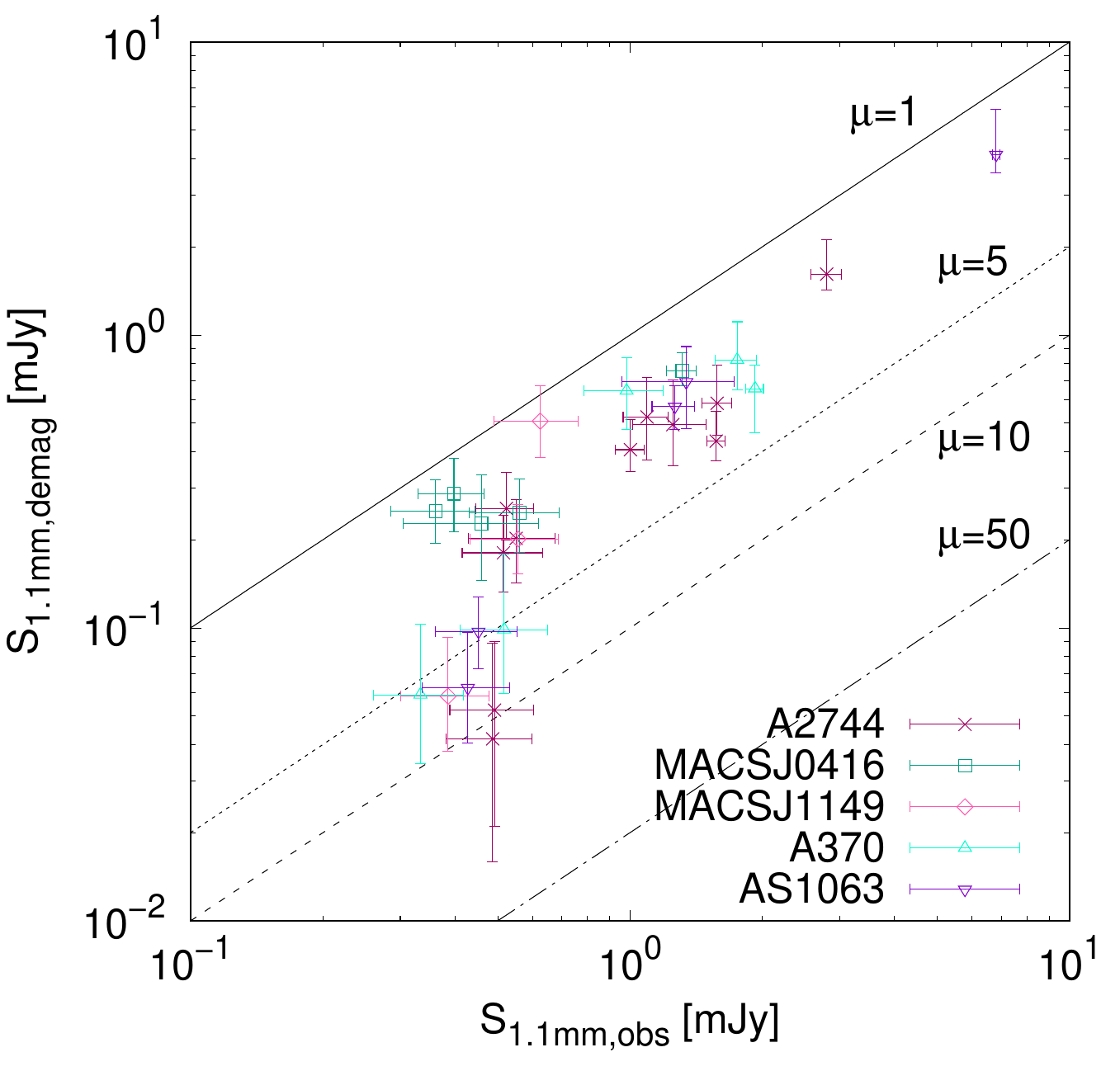}}
\caption{Median demagnified integrated flux density as a function of observed integrated flux density. Median values are obtained combining all models for each cluster field. Error bars in demagnified fluxes correspond to the 16th and 84th percentiles while for observed fluxes are $1\sigma$ statistical uncertainties. As a reference, black lines indicate magnification values of one (solid), five (dotted), ten (dashed) and 50 (dot-dashed).}
\label{fig_flux_intr_flux_obs}
\end{figure}

\begin{figure}
\centering
\resizebox{0.8\hsize}{!}
{\includegraphics{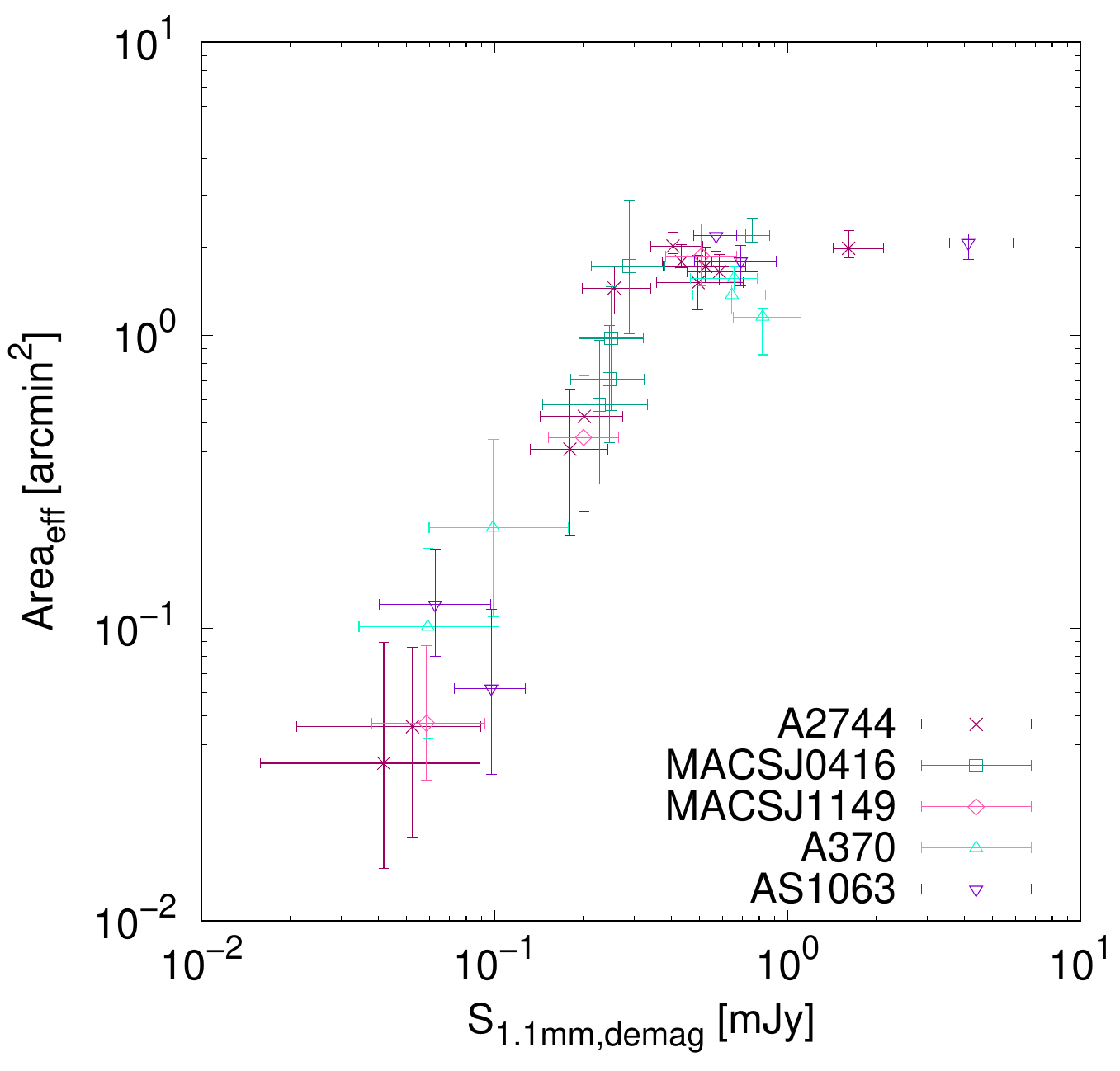}}
\caption{Median effective area as a function of demagnified integrated flux density. Median values are obtained combining all models for each cluster field. Error bars correspond to the 16th and 84th percentiles. For comparing uncertainty values, both axes cover the same interval in order of magnitude. Within the errors, both demagnified flux densities and effective areas span around 2.6 and 2.3 orders of magnitude, respectively.}
\label{fig_area_flux}
\end{figure}

We also use our Monte Carlo simulation realizations to obtain median estimates and uncertainties (from the 16th and 84th percentiles) in several source properties, as done in Paper IV. Source  magnifications, integrated flux densities (both observed and demagnified) and effective areas for the last two FFs are shown in Figs. \ref{fig_magnif_gal} to \ref{fig_area_flux}. Median (combined) magnification values for these clusters range from 1.3 to 11. Within the uncertainties, combined magnifications range from one to 29. For most sources, the discrepancies in the median values predicted by different models exceed their individual uncertainties, as Fig. \ref{fig_magnif_gal} shows. Median (combined) lensing-corrected flux densities range from $\sim0.04$ to $4.13\,\textmd{mJy}$. Within the uncertainties, combined demagnified flux densities cover around 2.6 orders of magnitude. In most of our cluster fields, our sources observed at low S/N have the largest magnifications, and thus the faintest observed sources are also the faintest intrinsically (see Fig. \ref{fig_flux_intr_flux_obs}). Median (combined) effective areas range from $\sim0.03$ to $2.19\,\textmd{arcmin}^2$. Within the uncertainties, combined effective areas cover around 2.3 orders of magnitude. Our faintest sources have the smallest effective areas (see Fig. \ref{fig_area_flux}).

\section{Results and discussion}\label{sect_results}

\begin{figure*}
\centering
\resizebox{0.7\hsize}{!}
{\includegraphics{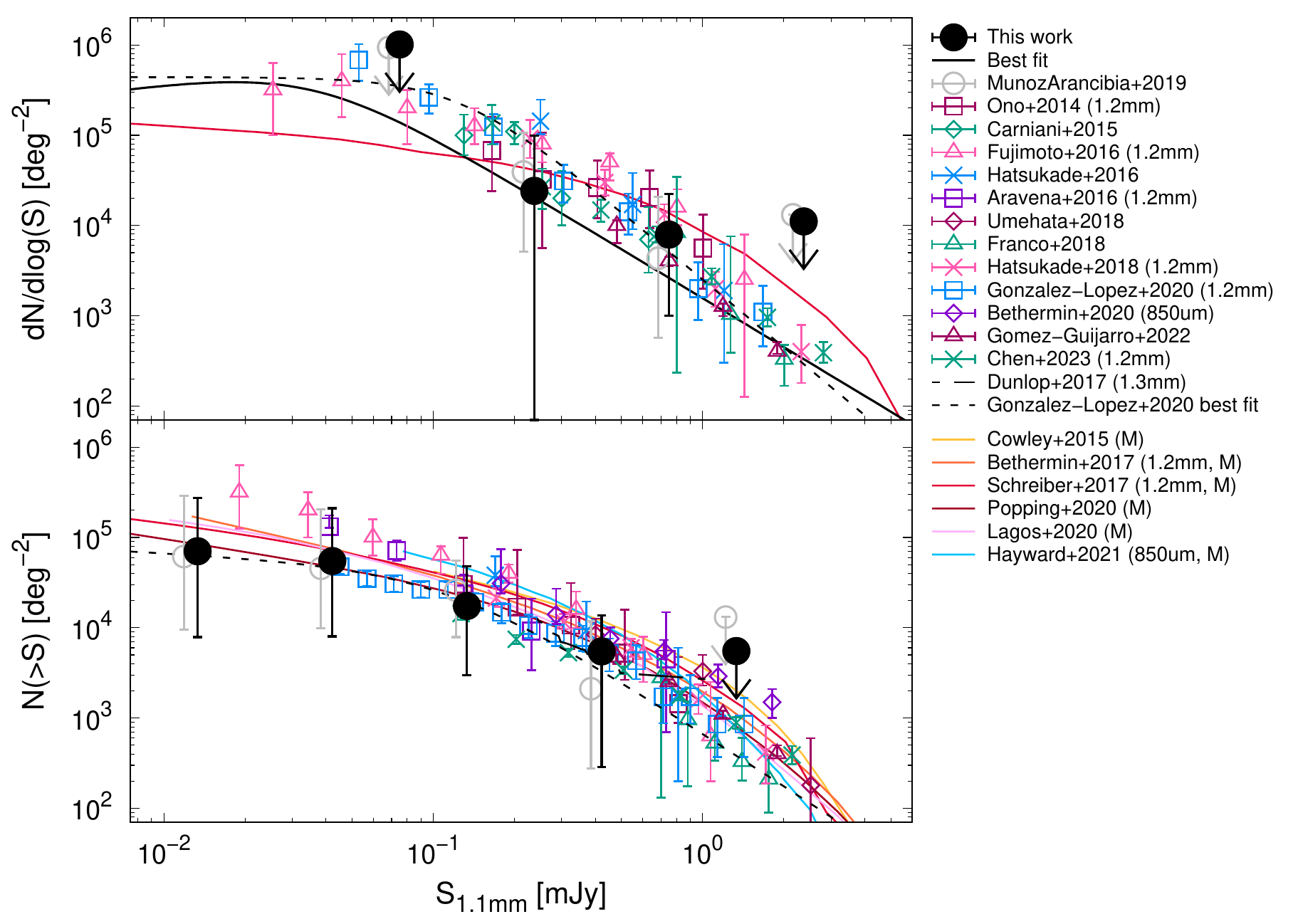}}
\caption{Differential (top) and cumulative (bottom) counts at $1.1\,\textmd{mm}$ compared to ALMA results and galaxy evolution model predictions from the literature. Our counts (large black filled circles) correspond to median values combining all models for all cluster fields. Error bars indicate the 16th and 84th percentiles, adding the scaled Poisson confidence levels for $1\sigma$ lower and upper limits respectively in quadrature. Arrows indicate $3\sigma$ upper limits for flux densities having zero median counts and non-zero values at the 84th percentile. We show our best double power-law fit as a black solid line (see $\S$\ref{sect_ebl}). We display the counts we derived for the first three FF clusters using $\textmd{PB}>0.5$ (\citealt{MunozArancibia2019}, Paper IV Corrigendum) as large gray open circles. We show previous results reported by \citet{Ono2014} as dark magenta squares, \citet{Carniani2015} as dark teal diamonds, \citet{Fujimoto2016} as magenta triangles, \citet{Hatsukade2016} as blue crosses, \citet{Aravena2016} as purple squares, \citet{Umehata2018} as dark magenta diamonds, \citet{Franco2018} as dark teal triangles, \citet{Hatsukade2018} as magenta crosses, \citet{Gonzalez-Lopez2020} as blue squares (with their best fit from a P(D) analysis shown as a black dashed line), \citet{Bethermin2020} as purple diamonds, \citet{Gomez-Guijarro2022} as dark magenta triangles, \citet{Chen2023} as dark teal crosses, and \citet{Dunlop2017} as a black dot-dashed curve. We show number counts predicted by the galaxy evolution models from \citet{Cowley2015} (yellow line), \citet{Bethermin2017} (orange line), \citet{Schreiber2017} (red line), \citet{Popping2020} (dark red line), \citet{Lagos2020} (light magenta line), and \citet{Hayward2021} (light blue line). We scale the counts derived at other wavelengths as $S_{1.1\,\textmd{mm}}=1.59\times S_{1.3\,\textmd{mm}}$, $S_{1.1\,\textmd{mm}}=1.27\times S_{1.2\,\textmd{mm}}$, and $S_{1.1\,\textmd{mm}}=0.51\times S_{850\,\mu\textmd{m}}$, assuming a modified black body (see $\S$\ref{sect_comp}).}
\label{fig_countsdiffcumu_comp}
\end{figure*}

\subsection{Number counts}\label{sect_counts}

We show our differential and cumulative number counts at $1.1\,\textmd{mm}$ in Fig. \ref{fig_countsdiffcumu_comp}, combining all models and cluster fields. Our counts combining models for each cluster field separately and altogether are presented in Table \ref{tab_counts} and Fig. \ref{fig_countsdiffcumu_field_tot}. We present counts down to the flux density bin centered on $0.024\,\textmd{mJy}$, as this is the faintest flux density bin where at least one cluster field has non-zero combined differential counts at the 84th percentile. We find variations across lens models in the median differential counts per flux bin up to $\approx0.6\,\textmd{dex}$ (although consistent within the error bars). Uncertainties coming from our Monte Carlo simulations differ by a factor of $\sim0.04-1.65$ ($\sim0.04-3.58$) from that predicted from Poisson statistics in the differential (cumulative) counts.

Combining all cluster fields, our differential counts span two orders of magnitude in demagnified flux density. However, at their limiting flux density bins (centered at $0.075\,\textmd{mJy}$ and $2.371\,\textmd{mJy}$) we report only $3\sigma$ upper limits, as their median counts are zero while having non-zero values at the 84th percentile. We note that at $0.024\,\textmd{mJy}$, only $\sim8\%$ of all realizations have non-zero counts, and thus combined differential counts are zero both in the median and 84th percentile (thus explaining the absence of a datapoint at $0.024\,\textmd{mJy}$ in Fig. \ref{fig_countsdiffcumu_comp} top panel). These realizations have a slight influence on the combined cumulative counts at $0.013\,\textmd{mJy}$, being $\approx0.1\,\textmd{dex}$ higher than those at $0.042\,\textmd{mJy}$.

At $0.13\,\textmd{mJy}$ and below, Monte Carlo simulation-based uncertainties dominate the combined cumulative counts over Poisson errors. The adoption of different values for the image-plane correction factors (i.e., completeness, flux deboosting and fraction of spurious sources) can change systematically the shape of the number counts as a whole; nevertheless, the main drivers of uncertainty in the faint-end number counts, for an individual lens model in the Monte Carlo simulations, are the uncertainties in the source-plane effective areas. For each ALMA detection, these are given by the uncertainties in source demagnified intensities (which in turn are given by those in both source magnification and observed intensity), but also by the steepness of the effective area vs demagnified peak intensity curve. The slope of this curve may vary with lens model, source redshift and demagnified peak intensity regime (see Paper IV). The steeper the curve, the larger the variation of $A_{\textmd{eff}}$ with $S_{\textmd{demag,peak}}$. The precision in both intensity measurements and source magnifications given by the lens models play an important role when applying our method for computing number counts.

\subsection{Cosmic variance}\label{sect_cosmicvar}

Number counts obtained in deep pencil beam surveys as the FF are limited by cosmic variance, i.e., the uncertainty in observational estimates of the number density of extragalactic objects due to the underlying large scale structure, which could become the dominant source of systematic error (e.g., \citealt{Somerville2004,Trenti2008,Moster2011}). We obtain $1\sigma$ cosmic variance estimates using three different approaches. First, we derive the field-to-field variance directly from our cumulative counts, computing the variance from the median counts per cluster field down to $0.013\,\textmd{mJy}$. Dividing its square root by the number of clusters and the corresponding median combined counts, we obtain a relative uncertainty of 17\%. Secondly, we obtain an approximate cosmic variance using the expression determined by \citet{Driver2010} from galaxies in the Sloan Digital Sky Survey Data Release 7 and generalized over all redshift. We adopt $2.28'\times2.28'$ as the (observed) survey area for each of our cluster fields, obtain comoving transverse lengths for these values at $z=2$, and use them in their Eq. 4 for one line of sight. We obtain a cosmic variance of 31\% (20\%) for the comoving volume contained in the range $z=1-3$ ($z=0-4$). Adopting a number of independent volumes equal to our number of cluster fields gives a cosmic variance of 14\% and 9\% for said redshift ranges. We note that \citet{Driver2010} assume the same cosmological parameters adopted in this work.

Finally, we obtain another estimate using the Cosmic Variance Calculator v1.03\footnote{\url{https://www.ph.unimelb.edu.au/~mtrenti/cvc/CosmicVariance.html}} by \citet{Trenti2008}, where cosmic variance is computed via the two point correlation function in extended Press-Schechter theory. Adopting a survey area $2.28'\times2.28'$, mean $z=2$, an intrinsic number of objects of four (as this is the median number of sources in our combined cumulative counts down to $0.013\,\textmd{mJy}$), a halo filling factor of 1.0, completeness 1.0, rms matter fluctuation today in linear theory
 $\sigma_8=0.9$ and a Sheth-Tormen bias, we obtain a relative cosmic variance of 18\% (14\%) for the redshift interval $z=1-3$ ($z=0-4$). Adopting a survey area $11.4'\times2.28'$ (i.e. as if our five cluster fields are displayed contiguously) and an intrinsic number of objects of 29 (as this is the total number of sources) gives a cosmic variance of 13\% and 10\% for said redshift ranges.. However, we note that some of the cosmological parameters used by this calculator are slightly different ($\Omega_m=0.26$, $\Omega_{\Lambda}=0.74$) to those adopted by us.

At our faintest flux limit, uncertainties in the median combined cumulative counts coming from our Monte Carlo simulations are in the range 75-281\%, while those from Poisson statistics are in the range 48-79\%. From these we note that cosmic variance estimates from the three approaches mentioned before are all exceeded by these uncertainties.

\citet{Popping2020} successfully reproduced the ASPECS-LP number counts \citep{Gonzalez-Lopez2020} using a semi-empirical model. They estimated the cosmic variance for the ASPECS-LP survey using 100 random sub-areas of their simulated sky, each covering an area the size of the ASPECS-LP region. They found that below $1\,\textmd{mJy}$, the typical $2\sigma$ scatter due to field-to-field variance is a factor of 1.5 at $1.1\,\textmd{mm}$, while sources brighter than $1\,\textmd{mJy}$ are typically missed by surveys of this angular size. Moreover, their model predicts that sources fainter than $1\,\textmd{mJy}$ have mostly stellar masses below $10^{11}\,\textmd{M}_{\odot}$, while sources brighter than $1\,\textmd{mJy}$ have mostly stellar masses above $5\times10^{10}\,\textmd{M}_{\odot}$. Our cumulative counts have a notable agreement with both the observed and modeled counts in ASPECS-LP (see $\S$\ref{sect_comp} and \ref{sect_flat}). Therefore, from their estimates we expect the bright end of our number counts ($>1\,\textmd{mJy}$) to be more affected by cosmic variance, missing a fraction of the objects with stellar mass in the range $5\times10^{10}-10^{11}\,\textmd{M}_{\odot}$ and most of the objects at $>10^{11}\,\textmd{M}_{\odot}$ that a wider-area survey may detect.

\subsection{Comparison to previous works}\label{sect_comp}

\begin{figure*}
\centering
\resizebox{0.65\hsize}{!}
{\includegraphics{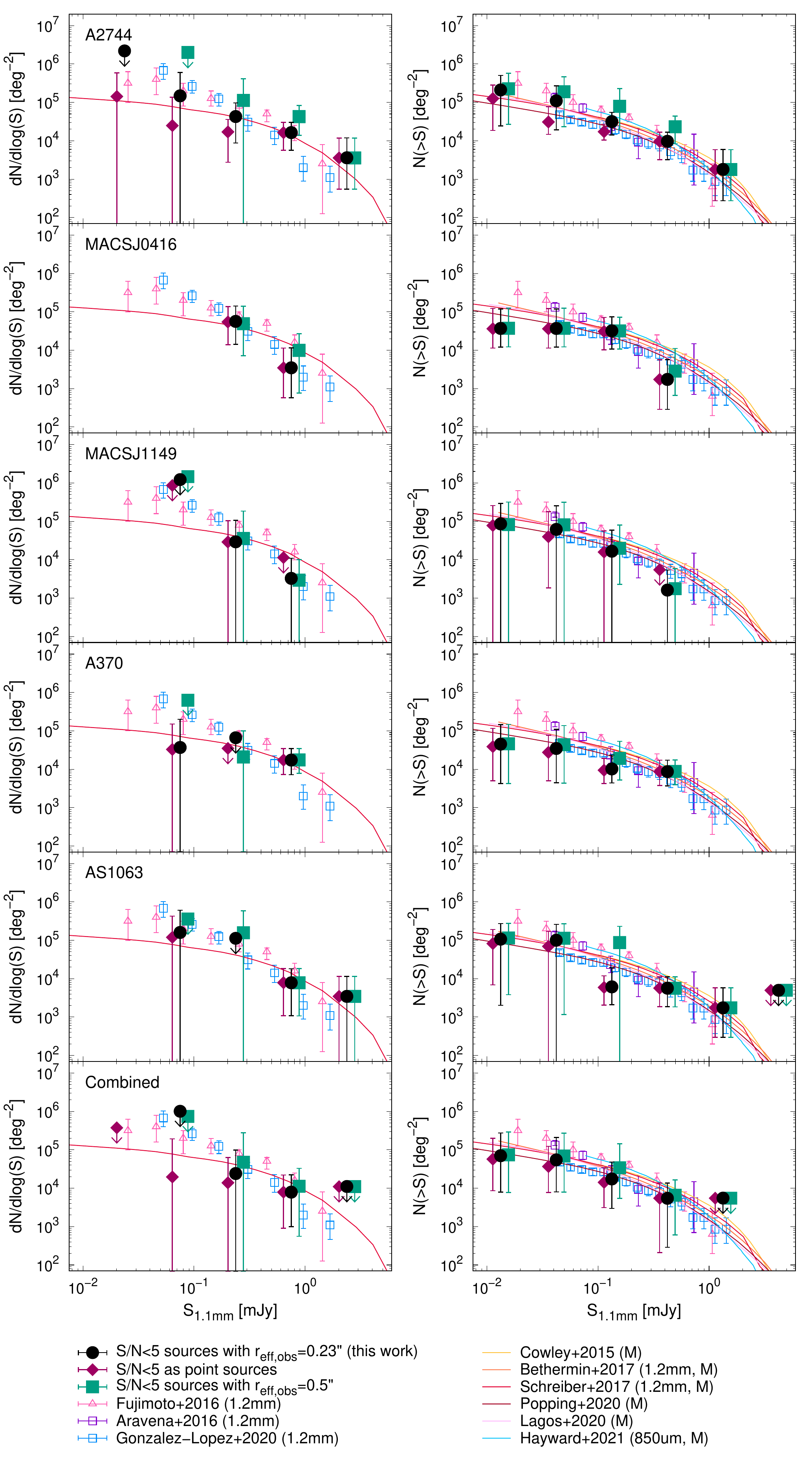}}
\caption{Differential (left) and cumulative (right) counts at $1.1\,\textmd{mm}$ for different assumptions regarding the image-plane source scale radii (half-light radii) for low-significance sources: adopting $r_{\textmd{eff,obs}}=0\farcs23$ (black filled circles, fiducial); assuming they are point sources (dark magenta filled diamonds); and adopting $r_{\textmd{eff,obs}}=0\farcs5$ (dark teal filled squares). Our counts correspond to median values. Error bars indicate the 16th and 84th percentiles, adding the scaled Poisson confidence levels for $1\sigma$ lower and upper limits respectively in quadrature. Arrows indicate $3\sigma$ upper limits for flux densities having zero median counts and non-zero values at the 84th percentile. We show previous results reported by \citet{Fujimoto2016} as magenta triangles, \citet{Aravena2016} as purple squares, and \citet{Gonzalez-Lopez2020} as blue squares. We show number counts predicted by the galaxy evolution models from \citet{Cowley2015} (yellow line), \citet{Bethermin2017} (orange line), \citet{Schreiber2017} (red line), \citet{Popping2020} (dark red line), \citet{Lagos2020} (light magenta line), and \citet{Hayward2021} (light blue line). We scale the counts derived at other wavelengths as in Fig. \ref{fig_countsdiffcumu_comp}.}
\label{fig_countsdiffcumu_reff}
\end{figure*}

We compare our results with the counts we derived using the first three FFs and adopting $\textmd{PB}>0.5$ (see Paper IV Corrigendum). Combining five FFs, using the $\textmd{PB}>0.3$ region and incorporating new spectroscopic redshifts for five sources in A2744, current counts are consistent to $1\sigma$ with our previous values, showing a similar flattening towards fainter flux densities. Our median differential (cumulative) counts change by a factor 0.6-1.8 (0.6-2.6). Uncertainties in the differential (cumulative) counts obtained from the Monte Carlo simulations change by a factor 0.6-4.0 (0.3-6.5), while scaled Poisson uncertainties change by a factor 0.8-1.4 (0.7-1.7). We note that in Paper IV, combined counts below $0.133\,\textmd{mJy}$ were contributed by two sources from A2744 and one from MACSJ1149. Now two sources from A370 and two from AS1063 are added to these flux density regime (see Fig. \ref{fig_area_flux}), and the flattening in the counts derived from them reinforces our previous results (see also Fig. \ref{fig_countsdiffcumu_reff}).

As in Paper IV, we explore the effect of adopting different source redshifts in the predicted counts, considering the following cases for all detections: a) assuming a Gaussian redshift distribution centered at $z=2\pm0.5$, b) adopting exactly $z=2$, c) assuming a uniform redshift distribution between the cluster redshift and $z=4$, and d) assuming a Gaussian redshift distribution centered at $z=3\pm0.5$. Combining all clusters, our fiducial counts are consistent with those obtained in all these cases, with variations in the median values up to $\approx0.1\,\textmd{dex}$. Case d), however, adds a $3\sigma$ upper limit at $0.024\,\textmd{mJy}$.

We compare our combined counts to results from recent ALMA observations that probe down to the sub-mJy level, as well as to predictions by state-of-the-art galaxy evolution models. To the literature data already compared in Paper IV at $1.1-1.3\,\textmd{mm}$ (for details see Paper IV and references therein), we add number counts derived from six new dedicated surveys and three new models: a $20\,\textmd{arcmin}^2$ survey at $1.1\,\textmd{mm}$ that contains a $z=3.09$ protocluster (ALMA deep field in SSA22, \citealt{Umehata2018}); a $69\,\textmd{arcmin}^2$ survey at $1.1\,\textmd{mm}$ in the deepest HST-WFC3 H-band part of the GOODS-South field (GOODS-ALMA, \citealt{Franco2018}); a $26\,\textmd{arcmin}^2$ survey at $1.2\,\textmd{mm}$ in the GOODS-South field (ASAGAO, \citealt{Hatsukade2018}); a $4.2\,\textmd{arcmin}^2$ survey at $1.2\,\textmd{mm}$ in the HUDF (ASPECS-LP, \citealt{Gonzalez-Lopez2020}); a $25\,\textmd{arcmin}^2$ serendipitous survey at $850\,\mu\textmd{m}$ around a sample of spectroscopically confirmed star-forming galaxies in the COSMOS and CDF-South fields (ALPINE, \citealt{Bethermin2020}); a $72\,\textmd{arcmin}^2$ survey at $1.1\,\textmd{mm}$ in the GOODS-South field, including both small and large spatial scales (GOODS-ALMA 2.0, \citealt{Gomez-Guijarro2022}); a $0.3\,\textmd{deg}^2$ serendipitous survey at $1.2\,\textmd{mm}$ towards 1001 ALMA calibrators (ALMACAL, \citealt{Chen2023}); a semi-empirical model for the dust continuum number counts of galaxies \citep{Popping2020}; a semi-analytic model of galaxy formation that includes Band 6 number counts predictions \citep{Lagos2020}; and a cosmological hydrodynamical simulation that predicts $850\,\mu\textmd{m}$ number counts \citep{Hayward2021}. Since we include in our comparison counts derived at wavelengths other than $1.1\,\textmd{mm}$, we scale their estimates as $S_{1.1\,\textmd{mm}}=1.59\times S_{1.3\,\textmd{mm}}$, $S_{1.1\,\textmd{mm}}=1.27\times S_{1.2\,\textmd{mm}}$, and $S_{1.1\,\textmd{mm}}=0.51\times S_{850\,\mu\textmd{m}}$. These factors are obtained assuming a modified black body model with a dust emissivity index $\beta=1.5$, dust temperature $T_{\textmd{dust}}=35\,\textmd{K}$, and $z=2.5$, which are values typically found for dusty galaxies (e.g. \citealt{Swinbank2014}).

Our combined counts are consistent within $1\sigma$ with most of the previous ALMA studies and galaxy evolution model predictions, shown in Fig. \ref{fig_countsdiffcumu_comp}. Moreover, our differential counts are consistent within $\approx1.6\sigma$ with \citet{Hatsukade2016} at $0.25\,\textmd{mJy}$, and our cumulative counts are consistent within $\approx1.1\sigma$ with \citet{Fujimoto2016} at $0.13\,\textmd{mJy}$. At $0.04\,\textmd{mJy}$, however, our median cumulative counts are lower than both \citet{Fujimoto2016} and \citet{Aravena2016} estimates by $\approx0.4\,\textmd{dex}$.

Extrapolating the cumulative counts of \citet{Fujimoto2016} to $0.01\,\textmd{mJy}$, our combined counts disagree beyond $1\sigma$. This difference may arise from a combination of several factors, including the assumed redshift distribution for sources contributing to counts at these flux densities (a Gaussian centered at $z=2\pm0.5$ by us vs adopting exactly $z=2.5$ by \citealt{Fujimoto2016}), depth in cluster ALMA data ($55-71\,\mu\textmd{Jy}\,\textmd{beam}^{-1}$ vs $38-41\,\mu\textmd{Jy}\,\textmd{beam}^{-1}$), and lens modeling uncertainties. We note that our cumulative counts in A2744 agree within $1\sigma$ when extrapolating their data to $0.01\,\textmd{mJy}$ (see Fig. \ref{fig_countsdiffcumu_reff}). Together with the different number of cluster fields (five vs one), this suggests that field-to-field variance may play a role in the difference with respect to our combined counts (see also $\S$\ref{sect_cosmicvar}).

Since we have assumed a given image-plane source size for low-significance detections (11/29 of our sample), we explore the effect of varying this source size by testing two extreme cases. As in Paper IV, these are: a) assuming these are point sources; and b) adopting an observed effective radius of $r_{\textmd{eff,obs}}=0.5''$, which gives integrated flux densities of $4.5\leq\textmd{S/N}<5$ sources scaling the peak intensities by the new ratios 0.46 and 0.35 in A370 and AS1063, respectively. These estimates for both differential and cumulative number counts are presented in Fig. \ref{fig_countsdiffcumu_reff}, for each cluster field separately and altogether, compared to our fiducial case.

When testing these two cases, median counts combining all clusters change up to $\approx0.3\,\textmd{dex}$ compared to the fiducial case, and are found to be consistent within $1\sigma$ with galaxy evolution models. Under the assumption that low-significance detections are point sources, median cumulative counts are consistent with \citet{Fujimoto2016} within $1\sigma$ at $0.42\,\textmd{mJy}$ and brighter, lower by $\approx0.6\,\textmd{dex}$ at $0.13\,\textmd{mJy}$ (consistent within $\approx1.9\sigma$), and lower by $\approx0.6\,\textmd{dex}$ at $0.04\,\textmd{mJy}$ (consistent within $\approx1.3\sigma$). They are consistent with \citet{Aravena2016} within $1\sigma$ down to $0.13\,\textmd{mJy}$ and within $\approx1.1\sigma$ at $0.04\,\textmd{mJy}$. They agree within $1\sigma$ with \citet{Gonzalez-Lopez2020}. Assuming $r_{\textmd{eff,obs}}=0.5''$ for $4.5\leq\textmd{S/N}<5$ sources, median cumulative counts agree within $1\sigma$ with \citet{Fujimoto2016}, \citet{Aravena2016} and \citet{Gonzalez-Lopez2020}.

\subsection{Number counts flattening}\label{sect_flat}

In line with results from Paper IV, our findings continue suggesting a flattening in the number counts below $\approx0.1\,\textmd{mJy}$. This flux density is consistent within $\approx2.6\sigma$ with the parameterisation we found in $\S$\ref{sect_ebl} ($S_0=3.6_{-2.3}^{+2.5}\times10^{-2}\,\textmd{mJy}$). \citet{Popping2020} predict the dust continuum flux density of model galaxies by coupling an observationally motivated model for their stellar mass and star formation rate (SFR) distributions with empirical scaling relations. Their model reproduces the flattening in the counts below $0.3\,\textmd{mJy}$ found by \citet{Gonzalez-Lopez2020}. We compare our results to \citet{Popping2020} cumulative number counts at $1.1\,\textmd{mm}$ for their entire simulated lightcone down to $\sim10^{-3}\,\textmd{mJy}$. We highlight the remarkable agreement between our median cumulative counts and the \citet{Gonzalez-Lopez2020} observational estimates (both their direct number counts and best fit from a P(D) analysis, and even extrapolating their cumulative counts to $0.01\,\textmd{mJy}$), as well as with the \citet{Popping2020} predictions.

\citet{Popping2020} also present model number counts decomposed in bins of redshift, stellar mass, SFR, and dust mass. In this scenario, main contributors to $1.1\,\textmd{mm}$ cumulative number counts in the range $0.01-0.1\,\textmd{mJy}$ are galaxies that lie at $z=1-2$; have stellar mass in the range $5\times10^9-10^{10}$ ($10^{10}-5\times10^{10}$) $\textmd{M}_{\odot}$ below (above) $0.03\,\textmd{mJy}$; SFR in the range $10-50\,\textmd{M}_{\odot}\,\textmd{yr}^{-1}$; and dust mass in the range $10^7-10^8$ ($10^8-10^9$) $\textmd{M}_{\odot}$ below (above) $0.08\,\textmd{mJy}$. Their predictions suggest that the number counts flattening is driven by the shape (specifically, the ``knee'' at $S_{1.1\,\textmd{mm}}\approx0.2-0.5\,\textmd{mJy}$, and shallow faint-end slope) of the (sub)mm luminosity function at $z=1-2$. We note, however, that their model does not account for gravitational lensing effects. Moreover, only one of the sources that contribute to our faint-end number counts has a reliable counterpart (AS1063-ID04, see Table \ref{tab_snr45gals} and Fig. \ref{fig_flux_intr_flux_obs}), preventing a detailed comparison with \citet{Popping2020} results.

Further characterization of the physical properties of faint-end sources in the five FFs (e.g. using SED fitting and/or models to estimate their photometric redshift, stellar mass and SFR, from ALMA data plus non-detections at other wavelengths) may help to constrain the validity of these predictions. This may also be aided via assembling a source catalog from a larger sample of cluster fields at mm wavelengths. Such a work in progress is the ALMA Lensing Cluster Survey (ALCS; Kohno et al., in prep.), as among its goals is deriving number counts towards 33 ALMA observed galaxy clusters (Fujimoto et al., in prep.).

\subsection{Extragalactic background light}\label{sect_ebl}

Using the Monte Carlo realizations of our differential number counts, we compute the contribution to the extragalactic background light (EBL) provided by each of them, by adding up the contribution contained in each flux bin. We find a median contribution of $14.3_{-7.4}^{+12.2}$ ($7.5_{-3.9}^{+6.4}$) $\textmd{Jy}\,\textmd{deg}^{-2}$ resolved in our demagnified sources at $1.1\,\textmd{mm}$ down to $0.013$ ($0.133$) $\textmd{mJy}$; here, uncertainties are computed from the 16th and 84th percentiles.

We compare our estimates to that by the Planck collaboration (using their best-fit extended halo model, see \citealt{Ade2014}) and a recent determination based on COBE/FIRAS and Planck observations \citep{Odegard2019}. In both cases, we first find their EBL values at $263.14\,\textmd{GHz}$ (the set Local Oscillator frequency for our observations) through interpolation of their data. At this frequency, we obtain EBL estimates of $19.1_{-0.7}^{+0.8}\,\textmd{Jy}\,\textmd{deg}^{-2}$ and $16.8_{-6.1}^{+9.6}\,\textmd{Jy}\,\textmd{deg}^{-2}$ for \citet{Ade2014} and \citet{Odegard2019}, respectively. The contribution provided by our demagnified sources at $1.1\,\textmd{mm}$ down to $0.013$ ($0.133$) $\textmd{mJy}$ represents $74.5_{-38.4}^{+64.0}\%$ ($38.9_{-20.1}^{+33.9}\%$) of the EBL by \citet{Ade2014} and $86.4_{-50.0}^{+120.4}\%$ ($45.6_{-26.2}^{+57.3}\%$) of that by \citet{Odegard2019}. We note that both EBL estimates chosen for comparison involved several assumptions, e.g. when calibrating the brightness scale and substracting the contribution of both cosmic microwave background and Galactic foreground emission, which may increase the uncertainties in the quoted EBL beyond reported values.

In order to estimate the EBL derived from our counts at $1.1\,\textmd{mm}$ for any flux density limit, we fit a double power-law function (e.g., \citealt{Scott2002}) to our differential number counts. This is given by
\begin{equation}
\phi(S)\textmd{d}S=\frac{N_0}{S_0}\left[\left(\frac{S}{S_0}\right)^{\alpha}+\left(\frac{S}{S_0}\right)^{\beta}\right]^{-1}\textmd{d}S, \label{eq_dpl}
\end{equation}
\noindent where $N_0$, $S_0$, $\alpha$, and $\beta$ are free parameters, and such that the integrated flux density down to a flux limit $S_{\textmd{lim}}$ can be computed as
\begin{equation}
S_{\textmd{integ}}=\int_{S_{\textmd{lim}}}^{\infty}S\phi(S)\textmd{d}S. \label{eq_dpl_ebl}
\end{equation}
\noindent The logarithmic form of Eq. \ref{eq_dpl} is given by
\begin{equation}
n(S)\textmd{d}\log(S)=\ln(10)\frac{N_0S}{S_0}\left[\left(\frac{S}{S_0}\right)^{\alpha}+\left(\frac{S}{S_0}\right)^{\beta}\right]^{-1}\textmd{d}\log(S), \label{eq_dpl_dndlogs}
\end{equation}
\noindent with $n(S)\textmd{d}\log(S)=\phi(S)\textmd{d}S$. We consider the ranges $N_0=10^3-10^6\,\textmd{mJy}\,\textmd{deg}^{-2}$, $S_0=0.01-1\,\textmd{mJy}$, $\alpha=0-5$, and $\beta=-1$ to 1. We fit Eq. \ref{eq_dpl_dndlogs} to each realization of our differential number counts, including only flux density bins that have non-zero values (thus allowing to extrapolate our counts to flux densities outside the regime covered by our ALMA sources).

From these fits, we find median parameters and associated uncertainties (from 16th and 84th percentiles) of $N_0=2.7_{-1.7}^{+5.0}\times10^5\,\textmd{mJy}\,\textmd{deg}^{-2}$, $S_0=3.6_{-2.3}^{+2.5}\times10^{-2}\,\textmd{mJy}$, $\alpha=2.8_{-0.8}^{+1.0}$, and $\beta=0.6_{-1.6}^{+0.4}$. Using Eq. \ref{eq_dpl_ebl} with the parameters found for each realization and $S_{\textmd{lim}}=0\,\textmd{mJy}$, we estimate a median total EBL associated with $1.1\,\textmd{mm}$ galaxies in the FFs of $21.2_{-9.6}^{+40.4}\,\textmd{Jy}\,\textmd{deg}^{-2}$. For $S_{\textmd{lim}}=0.013$ ($0.133$) $\textmd{mJy}$, we obtain a median contribution of $18.8_{-8.8}^{+40.4}$ ($5.0_{-3.1}^{+42.1}$) $\textmd{Jy}\,\textmd{deg}^{-2}$, corresponding to $94.2_{-14.8}^{+5.3}\%$ ($27.3_{-17.0}^{+63.0}\%$) of our estimate of the total EBL. This suggests that we may be resolving most of the EBL at $1.1\,\textmd{mm}$ down to $0.013\,\textmd{mJy}$.

\section{Concluding remarks}\label{sect_concl}

We have estimated lensing-corrected number counts at $1.1\,\textmd{mm}$. By adding $\textmd{S/N}\geq4.5$ ALMA detections in two FF galaxy clusters, A370 and AS1063, we have extended the study conducted in Paper IV with A2744, MACSJ0416 and MACSJ1149. The combination of these five FF clusters covers a total observed area of $\sim26\,\textmd{arcmin}^2$ ($\textmd{PB}>0.3$ region), which is reduced once lens models are applied (e.g., by $\sim2.7$ times after applying the CATS v4 model for a source-plane $z=2$).

Based on 29 ALMA detections over a $\textmd{PB}>0.3$ region, our counts combining all cluster fields span two orders of magnitude in demagnified flux density. They are consistent to $1\sigma$ with the counts we obtained for the first three FFs using the $\textmd{PB}>0.5$ region (see Paper IV Corrigendum). They also agree within $1\sigma$ with most of previous estimates using both ALMA observations and galaxy evolution model predictions. Below $\approx0.1\,\textmd{mJy}$, however, our cumulative number counts are lower by $\approx0.4\,\textmd{dex}$ compared to deep ALMA studies but ASPECS-LP, supporting the flattening in the counts we already reported in Paper IV. Also at this flux density and fainter, uncertainties coming from our Monte Carlo simulations dominate the cumulative counts over Poisson uncertainties.

We find systematic variations in the median differential counts across lens models up to $\approx0.6\,\textmd{dex}$, although being consistent within the error bars. For each individual lens model in our Monte Carlo simulations, the main drivers of uncertainty in the faint-end number counts are the uncertainties in the source-plane effective areas, dependent on source demagnified intensities. This highlights the importance, when applying our method, of having both precise intensity measurements and precise source magnifications given by the lens models.

Cosmic variance estimates are all exceeded by uncertainties in our median combined cumulative counts coming from both our Monte Carlo simulations and Poisson statistics. Using directly the Monte Carlo realizations of our differential number counts, we derive a median contribution to the EBL of $14.3_{-7.4}^{+12.2}$ ($7.5_{-3.9}^{+6.4}$) $\textmd{Jy}\,\textmd{deg}^{-2}$ resolved in our demagnified sources at $1.1\,\textmd{mm}$ down to $0.013$ ($0.133$) $\textmd{mJy}$. Using double power-law
function fits to these realizations instead, we obtain a median contribution of $18.8_{-8.8}^{+40.4}$ ($5.0_{-3.1}^{+42.1}$) $\textmd{Jy}\,\textmd{deg}^{-2}$; comparing this contribution to our estimate of the total EBL associated with $1.1\,\textmd{mm}$ galaxies in the FFs, suggests that we may be resolving most of the EBL at this wavelength down to $0.013\,\textmd{mJy}$.

Among current number counts estimates using ALMA data, our work is the one based on the largest set of ALMA-observed galaxy clusters reported to date. This number will increase once number counts from surveys like ALCS are released.

\begin{acknowledgements}

We gratefully acknowledge support from ANID through FONDECYT Postdoctoral Fellowship 3160776 (A.M.M.A.); FONDECYT Regular 1171710 (A.M.M.A., E.I.), 1221846 (E.I.), 1190818 (F.E.B.), 1200495 (F.E.B.), 1190335 (T.A.); Millennium Science Initiative Program - ICN12\_009 (F.E.B., T.A.), CATA-Basal - AFB-170002 (F.E.B., R.D.), ACE210002 (F.E.B., R.D.) and FB210003 (F.E.B., T.A., R.D.); Grant No. 2020750 from the United States-Israel Binational Science Foundation and Grant No. 2109066 from the United States National Science Foundation, and by the Ministry of Science \& Technology, Israel (A.Z.). A.M.M.A. acknowledges the AstroHackWeek 2018 event for helpful discussions that contributed to improve the code used in our analysis. The ALMA observations were carried out under programs ADS/JAO.ALMA\#2013.1.00999.S, ADS/JAO.ALMA\#2015.1.01425.S, and ADS/JAO.ALMA\#2017.1.01219.S. ALMA is a partnership of ESO (representing its member states), NSF (USA) and NINS (Japan), together with NRC (Canada) and NSC and ASIAA (Taiwan), in cooperation with the Republic of Chile. The Joint ALMA Observatory is operated by ESO, AUI/NRAO and NAOJ. This work is based on data and catalog products from HFF-DeepSpace, funded by the National Science Foundation and Space Telescope Science Institute (operated by the Association of Universities for Research in Astronomy, Inc., under NASA contract NAS5-26555). This work utilizes gravitational lensing models produced by PIs Brada{\v c}, Natarajan \& Kneib (CATS), Merten \& Zitrin, Sharon, Williams, Keeton, Bernstein and Diego, and the GLAFIC group. This lens modeling was partially funded by the HST Frontier Fields program conducted by STScI. STScI is operated by the Association of Universities for Research in Astronomy, Inc. under NASA contract NAS 5-26555. The lens models were obtained from the Mikulski Archive for Space Telescopes (MAST).

\end{acknowledgements}

\bibliography{biblio_aff2}

\begin{appendix}

\section{Number counts for each Frontier Fields galaxy cluster}

\begin{figure*}
\centering
\resizebox{0.65\hsize}{!}
{\includegraphics{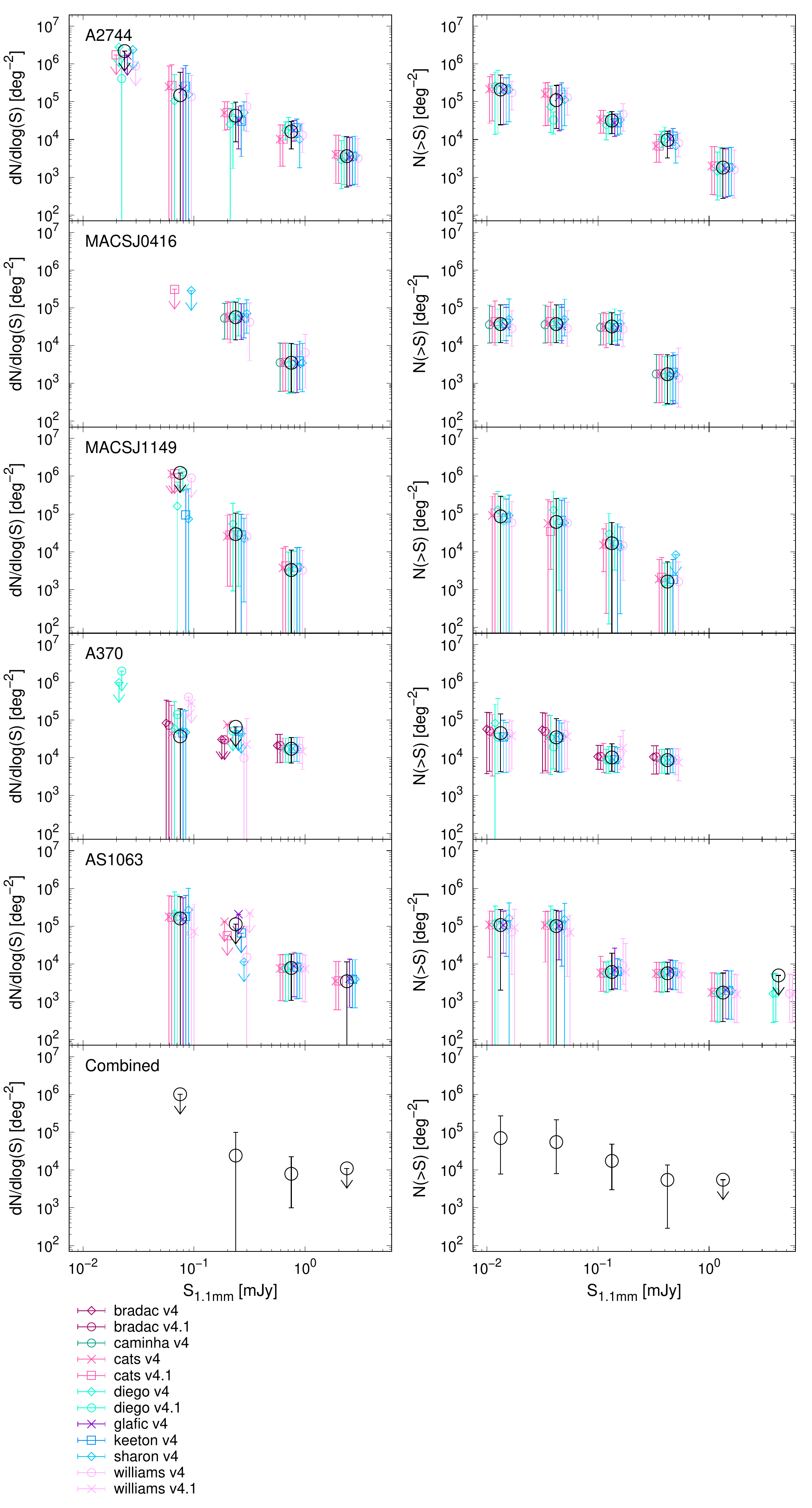}}
\caption{Differential (left) and cumulative (right) counts at $1.1\,\textmd{mm}$ (large black circles) for each cluster (see legends at top left) and combining all cluster fields (bottom panel). In the first five panels, we also show counts for each lens model (colored symbols) offset in flux around the combined counts for clarity. All values correspond to median counts. Error bars indicate the 16th and 84th percentiles, adding the scaled Poisson confidence levels for $1\sigma$ lower and upper limits respectively in quadrature. Arrows indicate $3\sigma$ upper limits for flux densities having zero median counts and non-zero values at the 84th percentile.}
\label{fig_countsdiffcumu_field_tot}
\end{figure*}

Here we present our differential and cumulative number counts at $1.1\,\textmd{mm}$, combining models for each cluster field separately and altogether in Table \ref{tab_counts}. These data are shown in Fig. \ref{fig_countsdiffcumu_field_tot}, together with number counts obtained for each lens model.

\begin{table*}
\begin{center}
\caption{Demagnified $1.1\,\textmd{mm}$ number counts.}
\begin{tabular}{c|ccc|ccc}
\hline \hline
Cluster field & $S_{1.1\,\textmd{mm}}$ & $\textmd{d}N/\textmd{d}\log(S)$ & \# sources & $S_{1.1\,\textmd{mm}}$ & $N(>S)$ & \# sources \\
& $(\textmd{mJy})$ & $(\textmd{deg}^{-2})$ & & $(\textmd{mJy})$ & $(\textmd{deg}^{-2})$ & \\
(1) & (2) & (3) & (4) & (5) & (6) & (7)\\
\hline
A2744 & 0.024 & $<2.164\times10^6$ & <3.0 & 0.013 & $(2.107_{-1.730}^{+2.788}\,_{-0.686}^{+0.957})\times10^5$ & $9.0_{-1.0}^{+1.0}\,_{-2.9}^{+4.1}$ \\
& 0.075 & $(1.479_{-1.479}^{+2.988}\,_{-1.221}^{+3.386})\times10^5$ & $1.0_{-1.0}^{+0.0}\,_{-0.8}^{+2.3}$ & 0.042 & $(1.098_{-0.830}^{+1.520}\,_{-0.358}^{+0.499})\times10^5$ & $9.0_{-1.0}^{+1.0}\,_{-2.9}^{+4.1}$ \\
& 0.237 & $(4.280_{-2.495}^{+3.489}\,_{-2.321}^{+4.143})\times10^4$ & $3.0_{-1.0}^{+3.0}\,_{-1.6}^{+2.9}$ & 0.133 & $(3.153_{-1.355}^{+1.808}\,_{-1.086}^{+1.547})\times10^4$ & $8.0_{-1.0}^{+1.0}\,_{-2.8}^{+3.9}$ \\
& 0.750 & $(1.633_{-0.725}^{+0.501}\,_{-0.779}^{+1.285})\times10^4$ & $4.0_{-2.0}^{+1.0}\,_{-1.9}^{+3.1}$ & 0.422 & $(9.681_{-4.922}^{+2.656}\,_{-4.165}^{+6.516})\times10^3$ & $5.0_{-3.0}^{+1.0}\,_{-2.2}^{+3.4}$ \\
& 2.371 & $(3.607_{-0.676}^{+0.336}\,_{-2.978}^{+8.254})\times10^3$ & $1.0_{0.0}^{+0.0}\,_{-0.8}^{+2.3}$ & 1.334 & $(1.800_{-0.339}^{+0.170}\,_{-1.486}^{+4.119})\times10^3$ & $1.0_{-0.0}^{+0.0}\,_{-0.8}^{+2.3}$ \\
& 7.499 & $0.000_{-0.000}^{+0.000}$ & $0.0_{-0.0}^{+0.0}\,_{-0.0}^{+1.8}$ & 4.217 & $0.000_{-0.000}^{+0.000}$ & $0.0_{-0.0}^{+0.0}\,_{-0.0}^{+1.8}$ \\
MACSJ0416 & 0.024 & $0.000_{-0.000}^{+0.000}$ & $0.0_{-0.0}^{+0.0}\,_{-0.0}^{+1.8}$ & 0.013 & $(3.711_{-1.769}^{+7.669}\,_{-1.769}^{+2.920})\times10^4$ & $4.0_{-1.0}^{+1.0}\,_{-1.9}^{+3.1}$ \\
& 0.075 & $0.000_{-0.000}^{+0.000}$ & $0.0_{-0.0}^{+0.0}\,_{-0.0}^{+1.8}$ & 0.042 & $(3.708_{-1.769}^{+7.659}\,_{-1.768}^{+2.918})\times10^4$ & $4.0_{-1.0}^{+1.0}\,_{-1.9}^{+3.1}$ \\
& 0.237 & $(5.645_{-2.903}^{+6.564}\,_{-3.062}^{+5.464})\times10^4$ & $3.0_{-1.0}^{+1.0}\,_{-1.6}^{+2.9}$ & 0.133 & $(3.212_{-1.500}^{+3.395}\,_{-1.531}^{+2.527})\times10^4$ & $4.0_{-1.0}^{+1.0}\,_{-1.9}^{+3.1}$ \\
& 0.750 & $(3.495_{-0.397}^{+0.757}\,_{-2.886}^{+7.999})\times10^3$ & $1.0_{-0.0}^{+0.0}\,_{-0.8}^{+2.3}$ & 0.422 & $(1.739_{-0.203}^{+0.265}\,_{-1.436}^{+3.980})\times10^3$ & $1.0_{-0.0}^{+0.0}\,_{-0.8}^{+2.3}$ \\
& 2.371 & $0.000_{-0.000}^{+0.000}$ & $0.0_{-0.0}^{+0.0}\,_{-0.0}^{+1.8}$ & 1.334 & $0.000_{-0.000}^{+0.000}$ & $0.0_{-0.0}^{+0.0}\,_{-0.0}^{+1.8}$ \\
& 7.499 & $0.000_{-0.000}^{+0.000}$ & $0.0_{-0.0}^{+0.0}\,_{-0.0}^{+1.8}$ & 4.217 & $0.000_{-0.000}^{+0.000}$ & $0.0_{-0.0}^{+0.0}\,_{-0.0}^{+1.8}$ \\
MACSJ1149 & 0.024 & $0.000_{-0.000}^{+0.000}$ & $0.0_{-0.0}^{+0.0}\,_{-0.0}^{+1.8}$ & 0.013 & $(8.612_{-7.227}^{+17.46}\,_{-5.546}^{+11.30})\times10^4$ & $2.0_{-1.0}^{+1.0}\,_{-1.3}^{+2.6}$ \\
& 0.075 & $<1.225\times10^6$ & <3.0 & 0.042 & $(6.162_{-4.872}^{+17.32}\,_{-3.968}^{+8.087})\times10^4$ & $2.0_{-1.0}^{+1.0}\,_{-1.3}^{+2.6}$ \\
& 0.237 & $(2.930_{-1.833}^{+3.733}\,_{-2.419}^{+6.704})\times10^4$ & $1.0_{-0.0}^{+0.0}\,_{-0.8}^{+2.3}$ & 0.133 & $(1.667_{-1.026}^{+1.872}\,_{-1.376}^{+3.815})\times10^4$ & $1.0_{-0.0}^{+1.0}\,_{-0.8}^{+2.3}$ \\
& 0.750 & $(3.274_{-3.274}^{+1.682}\,_{-2.703}^{+7.492})\times10^3$ & $1.0_{-1.0}^{+0.0}\,_{-0.8}^{+2.3}$ & 0.422 & $(1.617_{-1.617}^{+0.863}\,_{-1.335}^{+3.701})\times10^3$ & $1.0_{-1.0}^{+0.0}\,_{-0.8}^{+2.3}$ \\
& 2.371 & $0.000_{-0.000}^{+0.000}$ & $0.0_{-0.0}^{+0.0}\,_{-0.0}^{+1.8}$ & 1.334 & $0.000_{-0.000}^{+0.000}$ & $0.0_{-0.0}^{+0.0}\,_{-0.0}^{+1.8}$ \\
& 7.499 & $0.000_{-0.000}^{+0.000}$ & $0.0_{-0.0}^{+0.0}\,_{-0.0}^{+1.8}$ & 4.217 & $0.000_{-0.000}^{+0.000}$ & $0.0_{-0.0}^{+0.0}\,_{-0.0}^{+1.8}$ \\
A370 & 0.024 & $0.000_{-0.000}^{+0.000}$ & $0.0_{-0.0}^{+0.0}\,_{-0.0}^{+1.8}$ & 0.013 & $(4.517_{-3.478}^{+9.448}\,_{-2.154}^{+3.553})\times10^4$ & $4.0_{-1.0}^{+1.0}\,_{-1.9}^{+3.1}$ \\
& 0.075 & $(3.703_{-3.703}^{+14.00}\,_{-3.057}^{+8.474})\times10^4$ & $1.0_{-1.0}^{+0.0}\,_{-0.8}^{+2.3}$ & 0.042 & $(3.477_{-2.540}^{+6.935}\,_{-1.658}^{+2.735})\times10^4$ & $4.0_{-1.0}^{+1.0}\,_{-1.9}^{+3.1}$ \\
& 0.237 & $<6.615\times10^4$ & <3.0 & 0.133 & $(1.023_{-0.174}^{+0.885}\,_{-0.555}^{+0.990})\times10^4$ & $3.0_{-0.0}^{+1.0}\,_{-1.6}^{+2.9}$ \\
& 0.750 & $(1.735_{-0.378}^{+0.352}\,_{-0.941}^{+1.679})\times10^4$ & $3.0_{-1.0}^{+0.0}\,_{-1.6}^{+2.9}$ & 0.422 & $(8.693_{-1.590}^{+1.735}\,_{-4.715}^{+8.414})\times10^3$ & $3.0_{-1.0}^{+0.0}\,_{-1.6}^{+2.9}$ \\
& 2.371 & $0.000_{-0.000}^{+0.000}$ & $0.0_{-0.0}^{+0.0}\,_{-0.0}^{+1.8}$ & 1.334 & $0.000_{-0.000}^{+0.000}$ & $0.0_{-0.0}^{+0.0}\,_{-0.0}^{+1.8}$ \\
& 7.499 & $0.000_{-0.000}^{+0.000}$ & $0.0_{-0.0}^{+0.0}\,_{-0.0}^{+1.8}$ & 4.217 & $0.000_{-0.000}^{+0.000}$ & $0.0_{-0.0}^{+0.0}\,_{-0.0}^{+1.8}$ \\
AS1063 & 0.024 & $0.000_{-0.000}^{+0.000}$ & $0.0_{-0.0}^{+0.0}\,_{-0.0}^{+1.8}$ & 0.013 & $(1.072_{-0.920}^{+1.417}\,_{-0.511}^{+0.844})\times10^5$ & $4.0_{-1.0}^{+1.0}\,_{-1.9}^{+3.1}$ \\
& 0.075 & $(1.606_{-1.606}^{+2.568}\,_{-1.326}^{+3.675})\times10^5$ & $1.0_{-1.0}^{+1.0}\,_{-0.8}^{+2.3}$ & 0.042 & $(1.008_{-0.922}^{+1.392}\,_{-0.481}^{+0.793})\times10^5$ & $4.0_{-1.0}^{+1.0}\,_{-1.9}^{+3.1}$ \\
& 0.237 & $<1.126\times10^5$ & <3.0 & 0.133 & $(6.131_{-2.294}^{+11.80}\,_{-3.325}^{+5.935})\times10^3$ & $3.0_{-1.0}^{+0.0}\,_{-1.6}^{+2.9}$ \\
& 0.750 & $(7.844_{-4.488}^{+1.654}\,_{-5.051}^{+10.30})\times10^3$ & $2.0_{-1.0}^{+0.0}\,_{-1.3}^{+2.6}$ & 0.422 & $(5.654_{-2.242}^{+1.017}\,_{-3.067}^{+5.473})\times10^3$ & $3.0_{-1.0}^{+0.0}\,_{-1.6}^{+2.9}$ \\
& 2.371 & $(3.468_{-3.468}^{+0.506}\,_{-2.863}^{+7.937})\times10^3$ & $1.0_{-1.0}^{+0.0}\,_{-0.8}^{+2.3}$ & 1.334 & $(1.744_{-0.117}^{+0.244}\,_{-1.440}^{+3.991})\times10^3$ & $1.0_{-0.0}^{+0.0}\,_{-0.8}^{+2.3}$ \\
& 7.499 & $<9.954\times10^3$ & <3.0 & 4.217 & $<4.977\times10^3$ & <3.0 \\
Combined & 0.024 & $0.000_{-0.000}^{+0.000}$ & $0.0_{-0.0}^{+0.0}\,_{-0.0}^{+1.8}$ & 0.013 & $(7.026_{-5.263}^{+19.59}\,_{-3.350}^{+5.528})\times10^4$ & $4.0_{-1.0}^{+4.0}\,_{-1.9}^{+3.1}$ \\
& 0.075 & $<1.013\times10^6$ & <3.0 & 0.042 & $(5.486_{-3.883}^{+15.01}\,_{-2.616}^{+4.316})\times10^4$ & $4.0_{-1.0}^{+4.0}\,_{-1.9}^{+3.1}$ \\
& 0.237 & $(2.416_{-2.416}^{+5.012}\,_{-1.995}^{+5.529})\times10^4$ & $1.0_{-1.0}^{+2.0}\,_{-0.8}^{+2.3}$ & 0.133 & $(1.740_{-1.089}^{+2.579}\,_{-0.944}^{+1.685})\times10^4$ & $3.0_{-1.0}^{+4.0}\,_{-1.6}^{+2.9}$ \\
& 0.750 & $(7.929_{-4.688}^{+9.890}\,_{-5.106}^{+10.41})\times10^3$ & $2.0_{-1.0}^{+1.0}\,_{-1.3}^{+2.6}$ & 0.422 & $(5.470_{-3.800}^{+4.068}\,_{-3.523}^{+7.180})\times10^3$ & $2.0_{-1.0}^{+1.0}\,_{-1.3}^{+2.6}$ \\
& 2.371 & $<1.108\times10^4$ & <3.0 & 1.334 & $<5.526\times10^3$ & <3.0 \\
& 7.499 & $0.000_{-0.000}^{+0.000}$ & $0.0_{-0.0}^{+0.0}\,_{-0.0}^{+1.8}$ & 4.217 & $0.000_{-0.000}^{+0.000}$ & $0.0_{-0.0}^{+0.0}\,_{-0.0}^{+1.8}$ \\
\hline
\end{tabular}
\tablefoot{Column 1: cluster field. The bottom row group lists counts combining all cluster fields. Column 2: flux density bin for differential counts. Column 3: median differential counts per flux bin. For non-zero median counts, uncertainties are given separately using the 16th (84th) percentiles and scaled Poisson confidence levels for $1\sigma$ lower (upper) limits. For flux density bins having zero median counts and non-zero values at the 84th percentile, only $3\sigma$ upper limits are provided. Column 4: median number of sources per flux bin. Uncertainties are given separately using the 16th (84th) percentiles and Poisson confidence levels for $1\sigma$ lower (upper) limits. Column 5: flux density limit for cumulative counts. Column 6: median cumulative counts per flux limit. Uncertainties and upper limits are as in Column 3. Column 7: median number of sources per flux limit. Uncertainties are as in Column 4.}
\label{tab_counts}
\end{center}
\end{table*}

\end{appendix}

\end{document}